\documentclass[fleqn,usenatbib]{mnras}

\usepackage[T1]{fontenc}

\DeclareRobustCommand{\VAN}[3]{#2}
\let\VANthebibliography\thebibliography
\def\thebibliography{\DeclareRobustCommand{\VAN}[3]{##3}\VANthebibliography}

\usepackage{graphicx}	
\usepackage{amsmath}	
\usepackage{tabularx}
\usepackage{amssymb}	
\usepackage{orcidlink}

\usepackage{newtxtext,newtxmath}

\usepackage{xcolor}

\title[Decoding the spin-up epoch from Galactic bar]{Deciphering the Milky Way disc formation time encrypted in the bar chrono-kinematics}

\author[Zhang et al.]{Hanyuan Zhang$^{1}$\thanks{hz420@cam.ac.uk (hz420)}\orcidlink{0009-0005-6898-0927},
Vasily Belokurov$^{1}$\orcidlink{0000-0002-0038-9584}, N. Wyn Evans$^{1}$, Zhao-Yu Li$^{2,3}$\orcidlink{0000-0001-5017-7021}, Jason L. Sanders$^{4}$\orcidlink{0000-0003-4593-6788} and \newauthor Anke Ardern-Arentsen$^{1}$\orcidlink{0000-0002-0544-2217}
\\
$^{1}$ Institute of Astronomy, University of Cambridge, Madingley Road, Cambridge CB3 0HA, UK\\
$^{2}$ Department of Astronomy, School of Physics and Astronomy, 800 Dongchuan Road, Shanghai Jiao Tong University, Shanghai 200240, People's Republic \\ of China\\
$^{3}$ Key Laboratory for Particle Astrophysics and Cosmology (MOE) / Shanghai Key Laboratory for Particle Physics and Cosmology, Shanghai 200240, \\ People's Republic of China\\
$^{4}$ Department of Physics and Astronomy, University College London, London WC1E 6BT, UK
}

\date{Accepted XXX. Received YYY; in original form ZZZ}

\pubyear{2024}

\begin{document}
\label{firstpage}
\pagerange{\pageref{firstpage}--\pageref{lastpage}}
\maketitle

\begin{abstract}
We present a novel method to constrain the formation time of the Milky Way disc using the chrono-kinematic signatures of the inner Galaxy. We construct an O-rich Mira variable sample from the \textit{Gaia} Long-period Variable catalogue to study the kinematic behaviour of stars with different ages in the inner Galaxy.
From the Auriga suite of cosmological zoom-in simulations, we find that the age of the oldest stellar population with imprints of the bar in density and kinematics matches the disc spin-up epoch. This is because stars born before the spin-up show insufficient rotation and are not kinematically cold enough to be efficiently trapped by the bar.
We find that the bar kinematic signature disappears for Mira variables with a period shorter than 190 days. Using the period-age relation of Mira variables, we constrain the spin-up epoch of the Milky Way to be younger than $\sim11-12$~Gyr (redshift $\sim3$). We also discuss and compare our method and result to other evidence of the Milky Way spin-up epoch under the context of a realistic age uncertainty. Age uncertainty leads to an overestimation of the disc formation time when performing backward modelling. Our constrain of the spin-up epoch is independent from previous studies because it relies on the kinematics of the inner Galaxy instead of the solar vicinity.
\end{abstract}




\section{Introduction}

A disc is a typical component of many galaxies in the Universe. Disc formation and evolution in galaxies are studied with high-redshift observations of morphology \citep[e.g. ][]{Nelson_2023, Kartaltepe_2023, Robertson_2023} and gas kinematics \citep[e.g. ][]{Wisnioski_2019, Ubler_2019, deGraaff_2024}, as well as with cosmological simulations~\citep[e.g. ][]{Stern_2021, Hopkins_2023, Semenov_2024, Dillamore_2024}. In the Milky Way, because we can resolve individual stars and measure their complete 6D phase space positions, our Galaxy is an excellent laboratory to study disc formation and evolution~\citep[e.g. ][]{Carollo_2019, Sestito_2020, Di_Matteo_2020, Chandra_2023, Zhang_2024a, Nepal_2024, Gallart_2024}. 

The formation time of the Milky Way disc can be studied both using metallicity as a clock of Galactic evolution \citep{Norris_1985, Sestito_2020, Di_Matteo_2020, Belokurov_2022, Chandra_2023, Zhang_2024a, FernadezAlvar_2024, Viswanathan_2024} as well as using stellar ages directly \citep{Miglio_2021, Rix_2022, Xiang_2022, Queiroz_2023, Nepal_2024}. The metallicity distribution of the Galactic disc has received much attention in the past decades, especially at the metal-poor regime, which is potentially linked to the epochs of its early formation. Observations invoking the existence of a metal-weak thick disc are discussed periodically \citep[e.g. ][]{Norris_1985, Morrison_1990, Chiba_Beers_2000, Carollo_2010, Carollo_2019, An_Beers_2020, Di_Matteo_2020}. 

Recently, very metal-poor (VMP) stars, which are expected to be the oldest stars in the Milky Way, with disc-like orbits discovered by \citet{Sestito_2019, Sestito_2020} have revived the discussion. Although the disc-like VMP stars could have multiple origins ~\citep[see][]{Sestito2021}, the hypothesis that they come from an early proto-disc has attracted the most attention \citep[e.g. ][]{Di_Matteo_2020, Mardini_2022, Bellazzini_2024}. Using APOGEE data, \citet{Belokurov_2022} created a pure sample of in-situ MW stars and revealed that the net azimuthal velocity of the Galaxy increases abruptly across a short metallicity range around $\mathrm{[M/H]}\sim-1.3$. Comparing to a variety of numerical models of galaxy formation, \citet{Belokurov_2022} conclude that the fast observed {\it spin-up} of the Milky Way marks the emergence of the stable and dominant Galactic stellar disc. Note that while the metallicity scale of the Galactic spin-up is not very low, the corresponding lookback time is impressively large, corresponding to redshifts above $z>3$. Further support for this hypothesis has been found in independent studies of other simulation suites \citep[e.g.][]{Semenov_2024, Dillamore_2024, Chandra_2023}. Indeed, as these studies demostrate, our Galaxy was likely one of the first to form a stable stellar disc amongst objects of similar halo mass. With metallicities from the \textit{Gaia} BP/RP (XP) spectra \citep[e.g. ][]{Andrae_2023}, we can access an unprecedentedly large sample of metal-poor stars. Recent analyses of such {\it Gaia} XP-based samples bolster the idea that the Milky Way disc formed around $\mathrm{[M/H]}\sim-1.5$ to $-1$ \citep{Chandra_2023, Zhang_2024a, Kane2024}. Other origins for the VMP stars on disc-like orbits have also been discussed. These can form in the proto-Galaxy ~\citep{Sestito2021, Horta2024,Zhang_2024a}, or can be trapped by a decelerating bar~\citep{Dillamore2023, Li_2023, Yuan_2023}.

Although the stellar metallicity is usually measured more accurately than the age, the metallicity evolution also depends on the star formation history (sometimes leading to non-monotonic age-metallicity relation), which complicates the interpretation of results.
Accurate age measurements of disc stars can directly pin down the formation epoch of the Milky Way disc, but this is hard, and very few robust age estimates exist for low-metallicity stars. \citet{Miglio_2021} assessed the asteroseismological age of stars with light curves measured by the Kepler satellite and found the mean age of stars in the thick disc after deconvolving the age uncertainty is consistent with $\sim11$~Gyr. Similarly, \citet{Queiroz_2023} used the stellar ages from isochrone fitting and found the same age for the thick disc. Recently, \citet{Nepal_2024} found disc populations of stars older than $13$~Gyr using stellar parameters from the \textit{Gaia} RVS+XP \citep{Guiglion_2024}, which hints that the Galactic disc may have started to form within $1$~Gyr after the Big Bang. However, reconciling the early disc formation scenario with the cosmological simulations is challenging. In numerical models, the morphology of a galaxy appears to change quickly from spheroidal to disc-like when the galaxy reaches a critical mass threshold \citep{Hopkins_2023,Dillamore_2024}. Taking the diversity of the assembly histories into account, it appears unusual for a Milky Way-like galaxy to reach such a mass threshold much earlier than redshift $\sim3$ \citep{Belokurov_2022}. 

Previous measurements of the disc formation epoch use the kinematics of the solar vicinity or globular clusters. In this work, we provide a new independent method to measure the disc formation epoch that instead uses the chrono-kinematics of the Galactic bar.
The Galactic bar dominates the inner Galaxy and is identified from the stellar density distribution at the centre \citep{Nakada_1991, Stanek_1997, Wegg_2013}. Other than the spatial overdensity caused by the Galactic bar, it also induces distinct kinematic features in the inner Galaxy, which are useful when tracing the bar signatures. Due to the streaming motion of bar-supporting stars and their elongated orbits, the kinematics of the Galactic bar is characterised by the quadrupole radial velocity pattern \citep{Bovy_2019, Queiroz_2021, Hey_2023, Vislosky_2024, Liao_2024, Zhang_2024} and the bi-symmetric pattern in the fractional radial velocity map \citep{Zhang_2024}. 

The structure and formation of the bar has been explored intensively with N-body simulations \citep[e.g. ][]{Athanassoula_2002, Fragkoudi_2017, Fujii_2018, Ghosh_2024, Jimenez-Arranz2024}, and recently, with the Auriga cosmological simulation suite \citep{Grand_2017, Fragkoudi_2020, Merrow_2024, Fragkoudi_2024}. Several factors can impact the bar formation in a galaxy, such as disc kinematics \citep[e.g. ][]{Toomre_1964, Athanassoula_1986}, disc mass fraction and its assembly \citep[e.g. ][]{Fujii_2018, Bland-Hawthorn2023, Khoperskov_2024}, and gas fraction \citep[e.g. ][]{Athanassoula_2013}, but the exact conditions are still complicated \citep{Romeo_2023}. Many studies used N-body simulations to investigate the response of different stellar populations to the Galactic bar (e.g. classical bulge: \citealt{Saha_2012, Saha_2016}, the stellar halo: \citealt{Perez-Villegas_2017}, and the Galactic disc: \citealt{Debattista_2017, Fragkoudi_2017}). \citet{Debattista_2017} demonstrated that cold populations form a strong, thin, peanut-shaped bar, while hot populations form a weak, thick, boxy-shaped bar. As kinematically colder population respond to the Galactic bar more efficiently, the cold galactic disc provides the stars for the bar formation \citep[][]{Toomre_1964, Fragkoudi_2017, Debattista_2017, Boin_2024}. Moreover, early interactions with large satellites can seed bar formation. For example, \citet{Merrow_2024} show that an accretion event similar to the GS/E merger \citep[][]{Belokurov_2018, He18} in the MW can be responsible for the bar formation, as long as the GS/E progenitor is not too massive.

Long-period variables (LPV) are bright stars that follow tight period-luminosity relations \citep[][]{Rau_2019, Sanders_2023} and hence, are often used to trace the structure and kinematics of the inner Galaxy \citep[e.g. ][]{Catchpole_2016, Grady_2020, Semczuk_2022, Sanders_2022, Hey_2023, Zhang_2024}. Mira variables, a subcategory of LPVs, are high-amplitude thermally pulsating asymptotic giant branch stars. Their distances can be calibrated via the period-luminosity relation of O-rich Mira variables in the Milky Way \citep{Sanders_2023} and can have astrometry from \textit{Gaia} DR3. Mira variables also exhibit a period-age relation: Miras with shorter periods have older ages \citep{Feast_Whitelock_2000, Grady_2019, Trabucchi_2022, ZS23}. \citet{ZS23} re-calibrated the period-age relation of O-rich Mira variables to ensure consistency with the ages of Miras in globular clusters. Mira variables have been used to trace the chronological evolution of the Milky Way structures \citep[e.g. ][]{Catchpole_2016, Grady_2020, Semczuk_2022}, and recently, \citet{Sanders_2024} used the Mira variables in the nuclear stellar disc to constrain the bar formation epoch of the Milky Way.

In this paper, we use Mira stars to time the formation of the Galactic disc. We present the selection of O-rich Mira variable stars in the Milky Way and describe the properties of this sample in Section~\ref{sec::data}. In Section~\ref{sec::Auriga}, we briefly describe the Auriga cosmological simulations. The Auriga suite is used to study the correlation between the disc formation time and the kinematic signatures of the inner Galaxy in Section~\ref{sec::chrono_kinematic_simulation}. We build the connection between the "chrono-kinematic" signature of the Galactic bar and the spin-up time of the Galaxy in Section~\ref{sec::chrono_kinematic_simulation}. In Section~\ref{sec::spin_up_epoch_MW}, we use the Mira variable sample to study the kinematic signatures of populations with different ages in the inner Galaxy and to set a constraint on the disc formation epoch of the Milky Way. We compare and discuss our results with other measurements of the Milky Way disc formation time in Section~\ref{sec::discussion}. We also show that an age uncertainty would lead to an overestimation of the disc age when using backward modelling. Finally, in Section~\ref{sec::conclusion}, we summarise our methodology and conclusion. 

\section{Mira variables in Gaia LPV catalogue}
\label{sec::data}

We select O-rich Mira candidates from the Specific Object Study (SOS) table in the \textit{Gaia} DR3 LPV catalogue \citep{GaiaDR3_LPV}. The SOS table contains stars that are identified as LPV by the supervised classification pipeline in \cite{GaiaDR3_variable}, have 5th-95th percentile of their $G$ band measurements greater than 0.1 mag, and colour $G_{\mathrm{BP}}-G_{\mathrm{RP}} > 0.5$. The pulsation periods are determined using the generalised Lomb-Scargle periodogram \citep{VanderPlas_2018} from the $G$-band lightcurves spanning 34 months. The published amplitude, $\Delta G$, is the half peak-to-peak Fourier amplitude in the $G$ band. To complement {\it Gaia} data, we acquire infrared $JHK_s$ band photometry from the 2MASS survey \citep{2MASS}. There are $1\,657\,987$ LPV candidates after the cross-match between {\it Gaia} and 2MASS.

\begin{figure*}
    \centering
    \includegraphics[width=1.9\columnwidth]{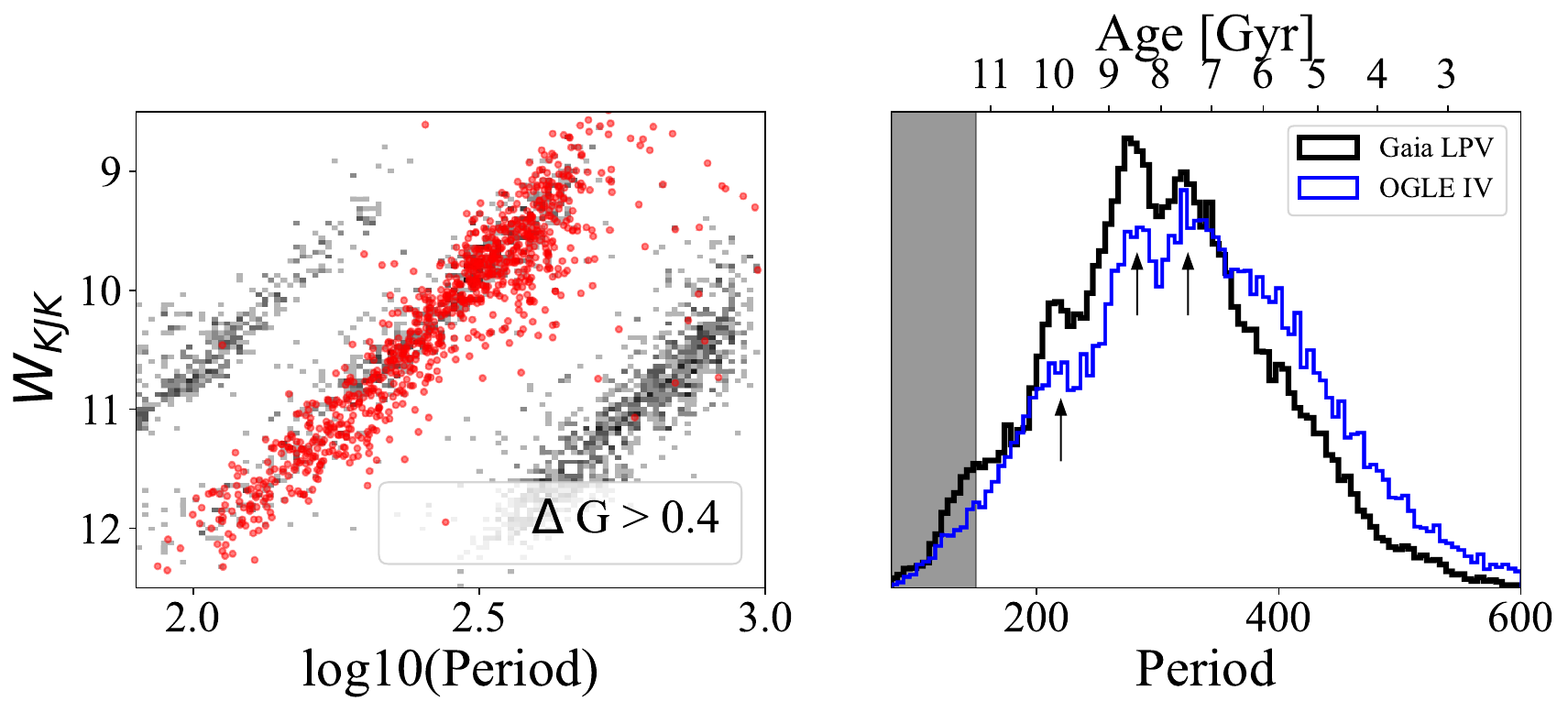}
    \caption{
    \textit{Left:} the period-luminosity plane for the LMC LPV candidates in the \textit{Gaia} DR3 LPV catalogue \citep{GaiaDR3_LPV}. LPV candidates with a high $G$-band amplitude are highlighted with red dots. The sequence occupied by the red dots are the period-luminosity relation of Mira variables. \textit{Right:} the period distribution of the final O-rich Mira variables in \textit{Gaia} LPV catalogue (black) and the O-rich Mira variable candidates in the OGLE survey (blue) \citep{Iwanek_2022}. The period is transformed to stellar ages using the period-age relation in \citet{ZS23}, and the ages are labelled on the top $x$-axis. Peaks in the period distribution are found at similar periods for both O-rich Mira variables in \textit{Gaia} LPV and OGLE catalogue and are highlighted by the black arrows.}
    \label{fig::amplitdue_cut_period_distribution}
\end{figure*}

\subsection{O-rich Mira variable selection}

Mira variables are characterised by their long pulsation period and large amplitude. Therefore, we select stars with a period between 80 to 1000 days \citep{Matsunaga_2009}, and remove stars with $\Delta G<0.4$ (compared to the cut at 0.43 in \citealt{Grady_2019}). To illustrate the performance of the amplitude cut, we present the period--luminosity plane for LPV candidates in the Large Magellanic Cloud (LMC) in the left panel of Fig.~\ref{fig::amplitdue_cut_period_distribution}. We further select LMC stars with $\Delta G>0.4$ and show their location in the period-luminosity plane with red dots. Most of the high amplitude stars occupy the middle period-luminosity sequence, the Mira variable sequence. Only $\sim5\%$ of high amplitude stars reside outside this middle sequence, which are the contamination in the Mira variable sample. We remove the bulk of contaminating young stellar objects by requiring $\texttt{best\_class\_score}>0.8$. 

Although Mira variables with a period shorter than $150$ days exist, we exclude LPV candidates with $\mathrm{period}<150$ days to remove potential short-period(SP)-red stars contamination. \citep[e.g.,][]{Feast_Whitelock_2000}. Kinematically, the SP-red stars behave similarly to longer-period Mira variables, which have a lower velocity dispersion. This indicates that rather than being older these stars are less dynamically heated and so have smaller ages.  A related result is found in \citet{ZS23}, in which the kinematics of the Mira sample with a period of 80 to 150 days is kinematically colder than expected. We omit the contaminating stars, many of which are likely SP-red objects from our analysis by neglecting Mira candidates with a period shorter than 150 days, which also removes the oldest Mira candidates ($\mathrm{age}\gtrsim11.5$~Gyr). 

Mira variables have oxygen(O)-rich or carbon(C)-rich chemistry. We focus on the O-rich Mira variables because they follow tighter period-luminosity relations \citep{Ita_Matsunaga_2011} and are more abundant in the Milky Way. O-rich Mira variables also have a better-calibrated period-age relation \citep{ZS23}. \citet{Sanders_Matsunaga_2023} provided an unsupervised classification of O- and C-rich Mira variable using the \textit{Gaia} XP spectrum, which performs better than the \textit{Gaia} classification for high extinction sources. We remove C-rich Mira variables as identified by \citet{Sanders_Matsunaga_2023} from our sample. When the XP spectrum is unavailable, we select O-rich Mira variables using periods, colours and amplitudes. \citet{Lebzelter_2018} found the $W_{\mathrm{BPRP}}-W_{\mathrm{JKs}}$ vs. $K_s$ plane can effectively separate O-rich and C-rich Mira variables in the LMC, where $W_{\mathrm{BPRP}}=G_{\mathrm{RP}}-1.3(G_{\mathrm{BP}}-G_{\mathrm{RP}})$ and $W_{\mathrm{JKs}}=K_s-0.686(J-K_s)$. \citet{ZS23} used period-amplitude and period-colour cuts to clean the C-rich Mira contamination. Here, we adopt similar cuts by selecting Mira candidates with $W_{\mathrm{BPRP}}-W_{\mathrm{JKs}}<1$, $\Delta G>1.2\log_{10}(\mathrm{period/days})-2.22$ and $G_{\mathrm{BP}}-G_{\mathrm{RP}}>7\log_{10} (\mathrm{period/days})- 13.20$.

There are in total $55\,586$ O-rich Mira variable candidates left after all the selections. The period distribution of the final sample is shown by the black histogram in the right panel of Fig.~\ref{fig::amplitdue_cut_period_distribution}. We see this distribution has three distinct peaks that are indicated by the black arrows. The locations of these period peaks do not agree with the expected aliasing frequency of the sparsely sampled \textit{Gaia} light curve \citep{GaiaDR3_LPV}. To demonstrate that these peaks are likely not arising from the sampling of the \textit{Gaia} light curves, we show the period distribution of the O-rich Mira candidates in the OGLE survey \citep{Iwanek_2022} as the blue histogram. The Mira candidates in \citet{Iwanek_2022} have a similar but not exactly the same coverage to our Mira variable candidates. The same peaks are found in the OGLE survey, in which photometric time series are densely sampled with less aliasing effects. Therefore, it is likely that these three peaks in the period distribution are physical. We hypothesise that these peaks may correspond to star-formation events in the Milky Way or, perhaps, stellar evolution effects. 

\subsection{Sample overview}
\subsubsection{Luminosity distance assignment}

\begin{figure*}
    \centering
    \includegraphics[width=1.9\columnwidth]{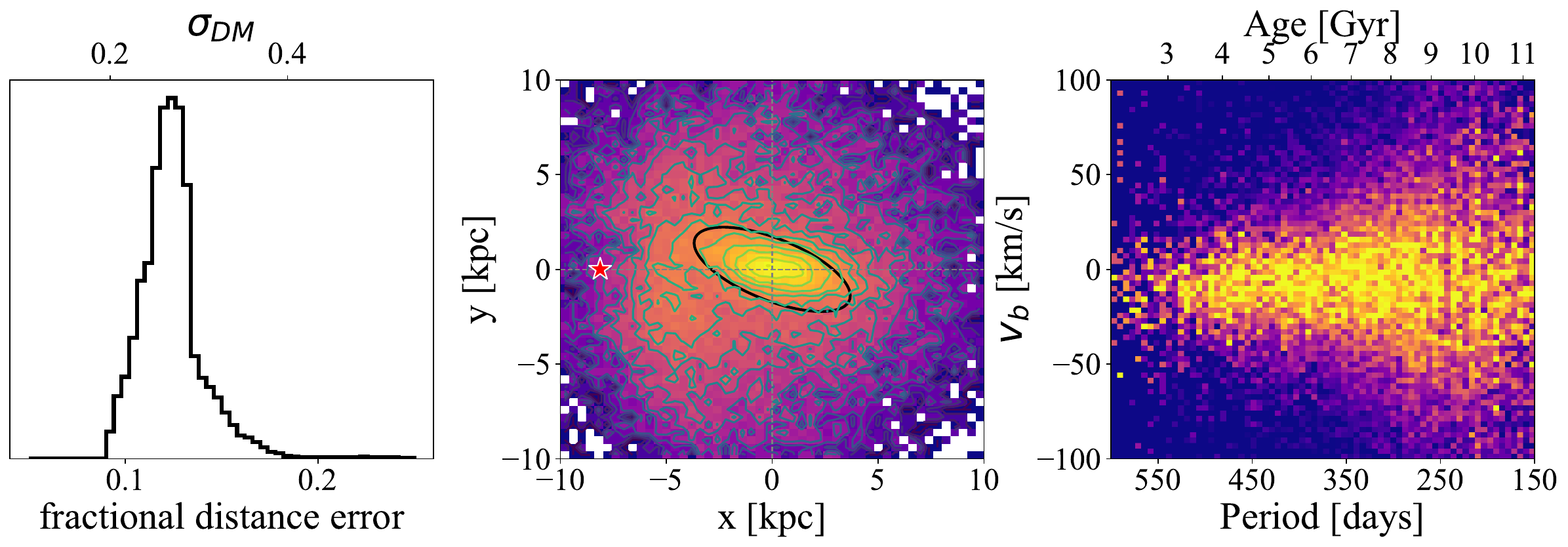}
    \caption{
    \textit{Left:} the distribution of the distance uncertainty (and the distance modulus uncertainty) of the final O-rich Mira variable sample. \textit{Middle:}  the distribution of Mira candidates in the Galactocentric $x-y$ plane. The white star in the middle panel labels the location of the Sun, and the black ellipse labels the Galactic bar. \textit{Right:} the column-normalised 2D histogram of the period vs. $v_b$ plane. The corresponding ages of Mira variables are assigned using the period-age relation in \citet{ZS23} and shown on the top $x$-axis.}
    \label{fig::disterr_xy_heating}
\end{figure*}

To calculate distance moduli, we use the period-luminosity relation for O-rich Mira variables in \citet{Sanders_2023}:
\begin{equation}
    M_{\mathrm{KJK}} = \left\{
    \begin{array}{rcl}
       -7.53 - 4.05(\log_{10}P-2.3),  & & \log_{10}P < 2.6,  \\
       -8.75 - 6.99(\log_{10}P-2.6) , & & \log_{10}P \geq 2.6,
    \end{array}
    \right.
    \label{plr}
\end{equation}
where $P$ is the period in days and $M_{\mathrm{KJK}}$ is the absolute, colour-corrected Wesenheit magnitude. The corresponding apparent Wesenheit magnitude is $W_{\mathrm{KJK}} = K_s - 0.473(J-K_s)$, and the distance modulus is then $\mathrm{DM} = W_{\mathrm{KJK}} - M_{\mathrm{KJK}}$. The extinction coefficient is taken from \citet{Wang_Chen_2019}. The scatter in the calibrated period-luminosity relation is 
\begin{equation}
    \sigma =  \left\{
    \begin{array}{rcl}
       \sigma_{23} + m_{\sigma_1} (\log_{10}P -2.3),  & & \log_{10}P < 2.6,  \\
       \sigma_{23} + 0.3 m_{\sigma_1} + m_{\sigma_2} (\log_{10}P-2.6),  & & \log_{10}P \geq 2.6,
    \end{array}
    \right.
\label{eqn::scatter_plr}
\end{equation}
where $\ln \sigma_{23} = -1.47$, $m_{\sigma_1} = 0.20$ and $m_{\sigma_2} = 0.89$. The scatter is contributed to by both the intrinsic dispersion in the period-luminosity relation and by the spread arising from using the single epoch 2MASS photometry. The final uncertainty in a distance modulus is the quadratic sum of the scatter in the period-luminosity relation, the uncertainty in the period determination and the photometric uncertainty. The distribution of the fractional distance uncertainty is shown in the left panel of Fig.~\ref{fig::disterr_xy_heating}. The median distance uncertainty in our sample is $\sim12\%$.

\subsubsection{Spatial distribution and the period-age correlation}

The spatial distribution of the sample in the Galactocentric $(x,y)$ plane is shown in the middle panel of Fig.~\ref{fig::disterr_xy_heating}. The Sun is placed at $x = -8.122\,\mathrm{kpc}$ \citep{GRAVITY_2018}. The sample has a good coverage of the Galactic disc on both the near and the far side. We can visually confirm the existence of the Galactic bar from the stretched and rotated overdensity in the centre. This detailed, panoramic view of the inner MW attests to the quality of our sample. 

To demonstrate the correlation between the pulsation period and the age of O-rich Mira variables, we present a column-normalized 2D histogram in the right panel of Fig.~\ref{fig::disterr_xy_heating}. This histogram shows the relationship between the pulsation period and the velocity in the Galactic latitude direction, denoted as $v_b$. The velocity is calculated as: $v_b = 4.74\mu_b\times d$ where $\mu_b$ is the proper motion in the Galactic latitude direction, and $d$ is the luminosity distance. It is important to note that this calculation of $v_b$ does not account for the solar reflex motion. As the right panel of Fig.~\ref{fig::disterr_xy_heating} clearly shows, the vertical velocity dispersion increases visibly with increasing (decreasing) Mira age (period). This is in line with well-established age--velocity dispersion relations \citep[][]{Stroemgren1946, Spitzer1951, Aumer_2016, Frankel2020, Sharma2021}.

\section{The Auriga simulation suite}
\label{sec::Auriga}

To be able to interpret the observed chrono-kinematic signatures presented here, we will compare our measurements to the behaviour of stellar populations in numerical simulations of MW-like galaxies. More precisely, we explore the bar signatures of different mono-age populations using the Auriga simulations \citep{Grand_2017, Grand_24}. The Auriga simulations are a suite of magneto-hydrodynamical zoom-in simulations for galaxies that cover a large mass range (virial mass between $M_{200}=0.5-2\times10^{12}\,M_{\sun}$ at $\tilde{z}=0$, where $\tilde{z}$ is the redshift). The Auriga simulations run from $\tilde{z} = 127$ to $\tilde{z} = 0$ with cosmological parameters $\Omega_m = 0.307$, $\Omega_\Lambda = 0.693$, $\Omega_b = 0.048$ and a Hubble constant of $h = 0.677$ \citep[][]{Plank_2014}. The Auriga simulation suite has a comprehensive galaxy formation model that includes the primordial and metal-line cooling, a uniform ultraviolet background for reionisation that completes at $\tilde{z}=6$, a multi-phase subgrid model for the interstellar medium, star formation, stellar evolution and feedback, magnetic field, black hole seeding and accretion, and AGN feedback \citep[see more detail in][]{Grand_2017}. The simulations are run using the magnetohydrodynamic simulation code \textsc{AREPO} \citep{Springel_2010, Pakmor_2016}.

The Auriga simulations have good mass resolution: the dark matter particles have masses $\sim4\times10^{5}\,M_{\sun}$, and the gas and star particles have masses $\sim5\times10^{4}\,M_{\sun}$. The collisionless dark matter and star particles have a softening length of $500h^{-1}$ pc for $\tilde{z}>1$ and 369 pc later. Auriga galaxies show structural and kinematics properties similar to observed disc galaxies. In particular, they have long-lived galactic bars and boxy/peanut bulges formed from buckling instabilities \citep{Fragkoudi_2020}. Therefore, galaxies in the Auriga simulations are good laboratories to investigate the chrono-chemo-kinematic evolution of galaxies and the Galactic bar \citep{Fragkoudi_2020, Fragkoudi_2024}.

In this paper, we present the analysis and results of three Milky Way-like, barred Auriga galaxies (Au18, Au23 and Au26). These three halos are also analysed in \citet{Fragkoudi_2020} and are found to be viable Milky Way analogues. In particular, \citet{Fattahi_2019} showed that the halo of Au18 has high radial anisotropy, which is caused by a Gaia-Sausage-Enceladus(GSE)-like \citep{Belokurov_2018,He18} merger event that happened 9 Gyr ago (in Au18). \citet{Merrow_2024} demonstrated that the bar formation in Au18 is triggered by tidal forces induced during the GSE-like merger. Therefore, Au18 provides a valuable insight into the bar formation and evolution of the Milky Way due to its similar assembly history \citep{Fattahi_2019, Merrow_2024}. However, finding perfect surrogates of the Milky Way from cosmological simulations would be tricky. Hence, we only use Auriga galaxies as laboratories to test the methodology in this work.

\begin{figure*}
    \centering
    \includegraphics[width=0.99\columnwidth]{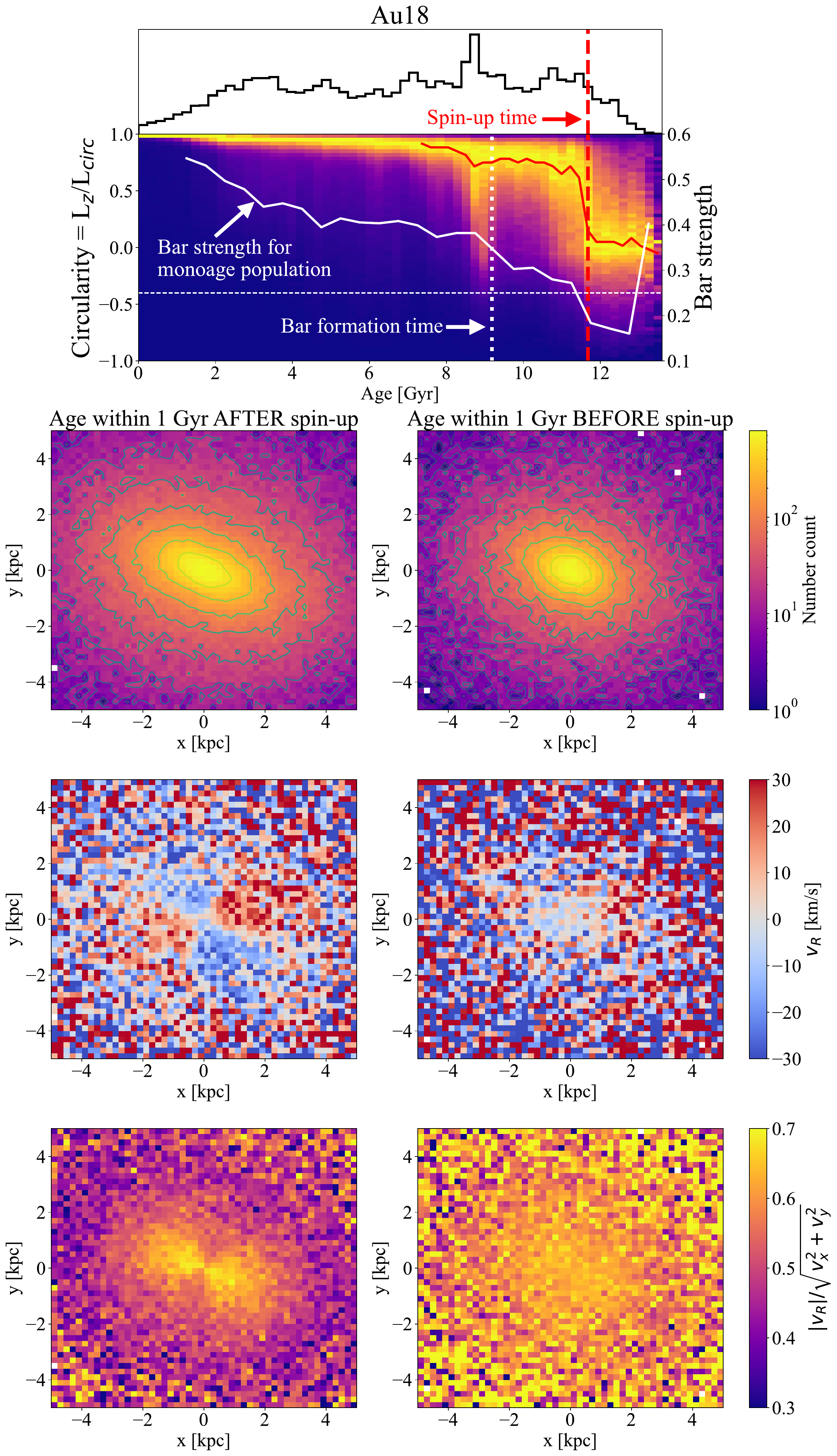}
    \includegraphics[width=0.99\columnwidth]{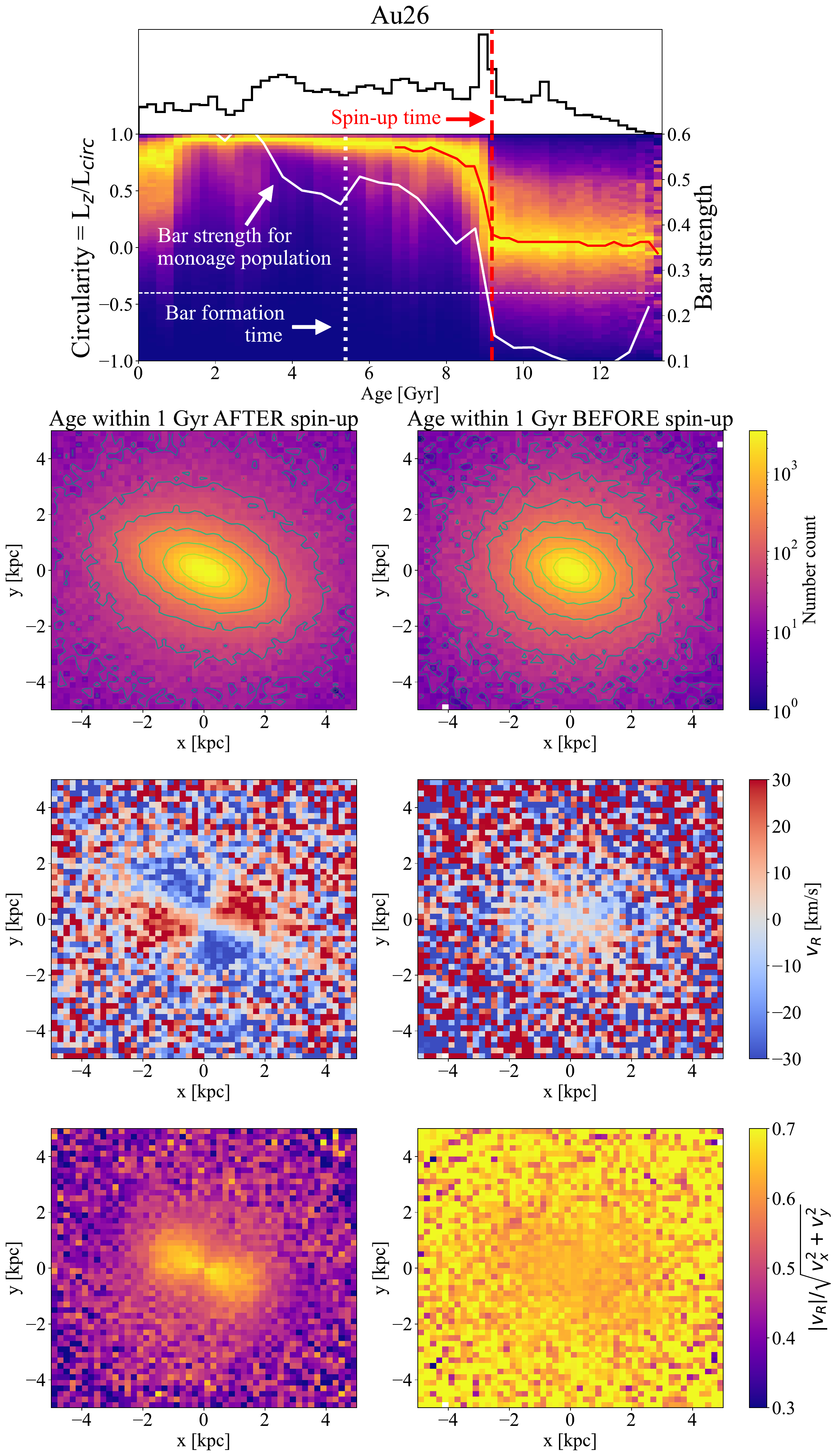}
    \caption{A demonstration that the bar signatures appears in populations born after the spin-up but not before using Au18 (left) and Au26 (right). \textit{Top row:} the age distribution and the column-normalised 2D histogram of the age-circularity plane. The white-solid lines inside the age-circularity diagrams are the bar strength defined by Eq.~\ref{eqn::bar_strength} for populations of different ages. The horizontal white dashed line is the strong bar criterion chosen in \citet{Fragkoudi_2024}, and the vertical white dotted line is the bar formation time in these two galaxies reported in \citet{Fragkoudi_2024}. The red-solid line denotes the circularity for each mono-age population, and the red-dashed line labels the spin-up time. In the lower sets of panels, we show stars born after 1 Gyr of the spin-up in the left column, and stars born before 1 Gyr of the spin-up in the right column. The first row of plots in the lower section shows the surface density distribution; the second row plots the mean radial velocity field, and the third row shows the $|v_R|/v_\mathrm{tot}$ map, which $v_\mathrm{tot}$ is the total in-plane velocity.}
    \label{fig::Au1826_bar_appearing}
\end{figure*}

\begin{figure*}
    \centering
    \includegraphics[width=1.98\columnwidth]{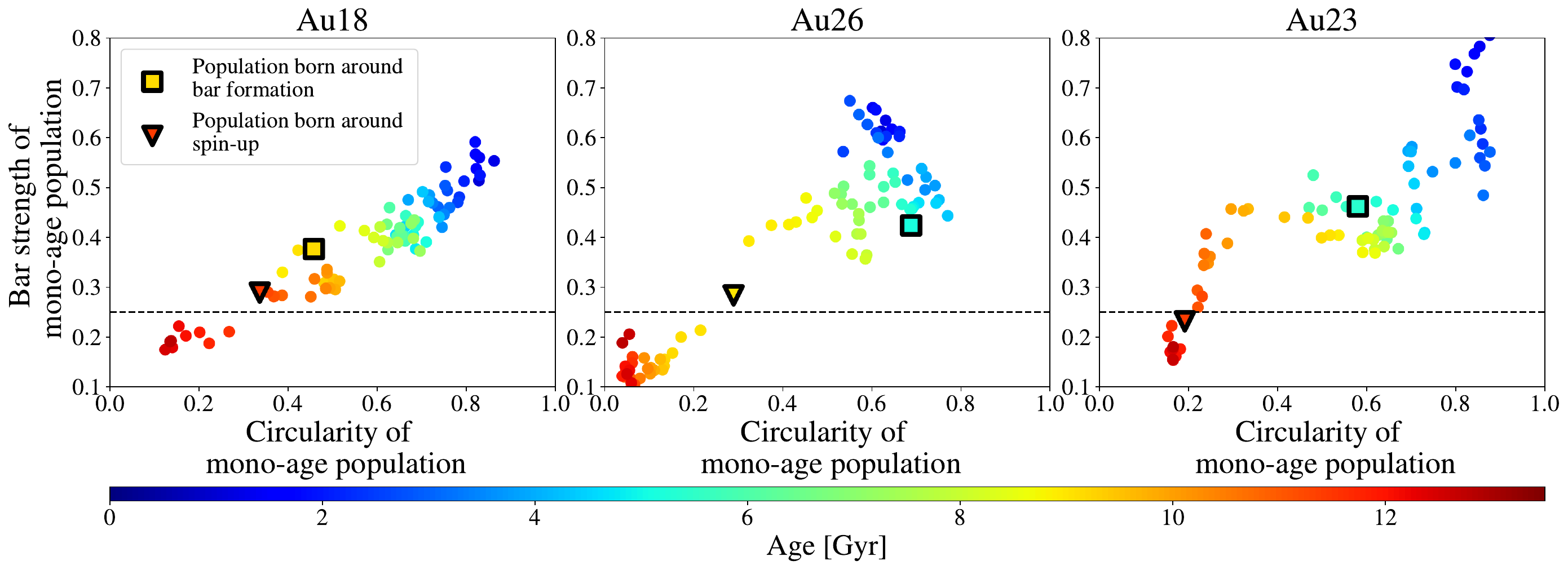}
    \caption{Circularity vs. bar strength for mono-age populations in Au18, 26 and 23. The colour of each dot represents the average age of the mono-age population. Each mono-age population are binned with 0.5 Gyr age segments. The black dashed lines label the bar strength threshold at $A_2 = 0.25$. The square/triangle with black edges labels the circularity and bar strength for the population that born at the moment of bar formation/spin-up in each Auriga galaxy.}
    \label{fig::circularity_vs_A2}
\end{figure*}

\begin{figure*}
    \centering
    \includegraphics[width=0.9\textwidth]{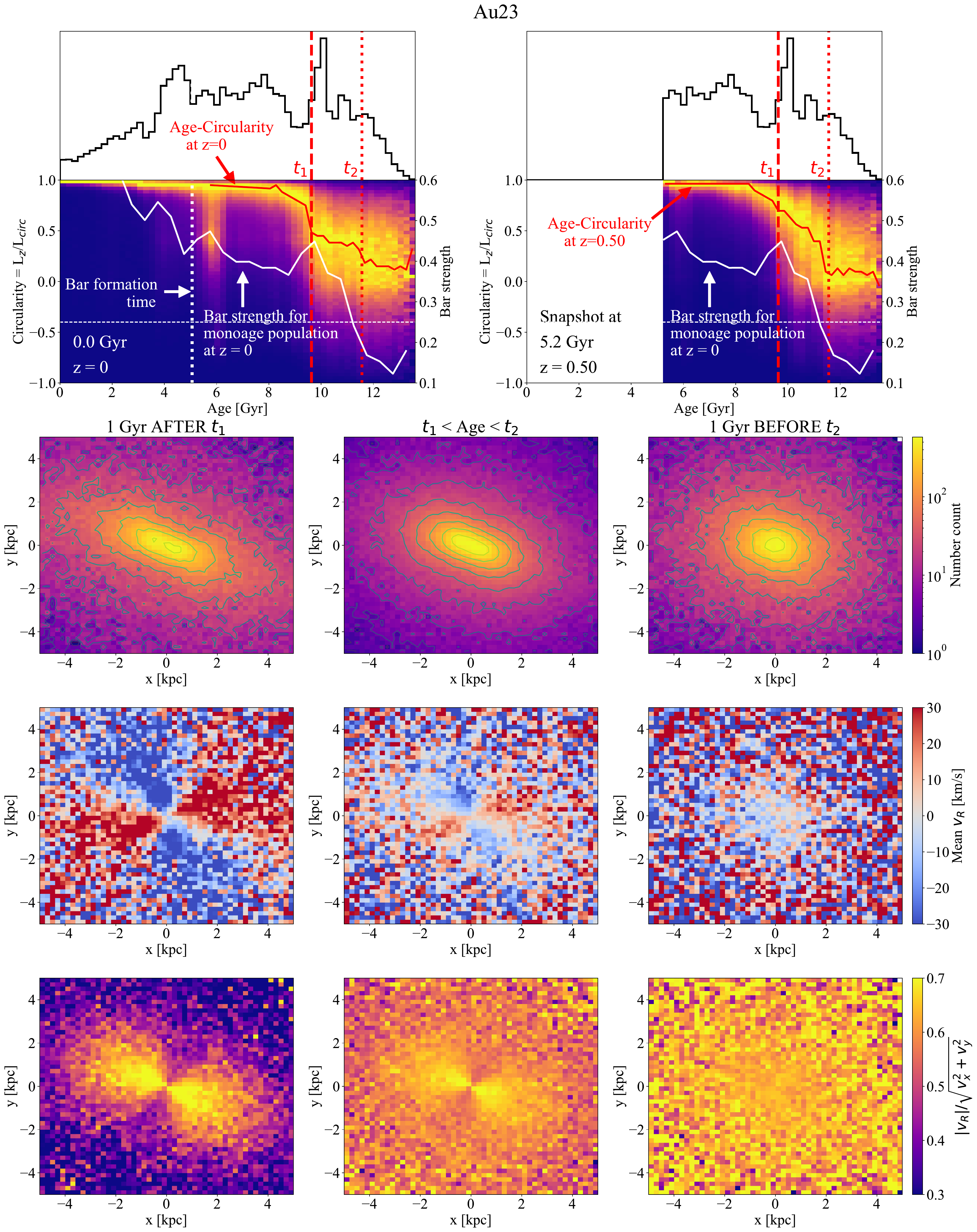}
    \caption{Similar to Fig.~\ref{fig::Au1826_bar_appearing} but for Au23. 
    \textit{Top left:} the column-normalised age-circularity plane for Au23 at $\tilde{z}=0$. The red-dashed and red-dotted lines are used to label two increments in circularity at $t_1$ and $t_2$ in Au23, seen at $\tilde{z}=0$. The red solid line show the trends of the circularity as a function of ages. The white solid line is the bar strength of different mono-age population observed at $\tilde{z}=0$. The vertical white dashed line label the bar formation time in Au23 \citep{Fragkoudi_2024}. \textit{Top right:} the age-circularity plot but for Au23 at $\tilde{z}=0.504$ ($t_{\mathrm{lookback}}\sim5.2$~Gyr). The vertical red dashed lines are the same as the top left panel that labels the ages of increment in circularity in the $\tilde{z}=0$ snapshot. The white solid line is also the same that shows the bar strength for different mono-age population in the $\tilde{z}=0$ snapshot. The red solid lines shows the age--circularity trends in $\tilde{z}=0.504$ snapshot, before the moment of bar formation. In the lower set of panels, we select stars born within 1 Gyr after $t_1$, between $t_1$ and $t_2$, and within 1 Gyr after $t_2$ and show their present-day ($\tilde{z}=0$) spatial and kinematic distribution in the left, middle and right columns, respectively. The {\it top row} in the lower sets is the spatial density distribution. The {\it middle row} is the radial velocity field. The {\it bottom row} is the $|v_R|/v_\mathrm{tot}$ map.}
    \label{fig::Au23}
\end{figure*}

\section{Chrono-kinematics of the galactic bar linked to the disc formation}
\label{sec::chrono_kinematic_simulation}

Hints on the bar-disc correlation in the Milky Way can be found in the chemo-kinematics of the Galactic bar and disc. Several recent studies focused on the chemo-kinematics in the solar vicinity have shown that the Milky Way does not host a significant disc population with metallicity below $\mathrm{[M/H]}\lesssim-1.5\sim-1.0$ compared to the halo contribution at the same metallicity \citep{Belokurov_2022, Chandra_2023, Zhang_2024a}. Observations of the inner Galaxy show that the bar stars rarely have metallicity below~$-1$ \citep{Arentsen_2020,Queiroz_2021}. Recently, \citet{Liao_2024} revealed the chemo-kinematic pattern of the near side of the Galactic bar, and showed that the kinematic signature of the Galactic bar disappears below $\mathrm{[M/H]}\lesssim-1$. The lower bound of the metallicity of the Galactic bar is similar to the metallicity when the Milky Way begins to spin up, which hints at the possibility of using the bar kinematics to set constraints on the spin-up time of the Milky Way. 

Irrespective of the mechanism seeding the Galactic bar, kinematically colder stars in the disc always respond more efficiently to its presence ~\citep[e.g.][]{LB79,CDE,Fragkoudi_2017, Debattista_2017,Boin_2024}.
In the Milky Way, stars born before the disc formation are hypothesised to belong to the Aurora/proto-Galaxy population with halo-like kinematics \citep{Belokurov_2022, Rix_2022}, although some net rotation may exist \citep{Chandra_2023, Zhang_2024a, Arentsen_2024}. As halo stars are kinematically hotter and have significantly lower rotation compared to disc stars \citep{Belokurov_2022, Rix_2022}, stars born in the galactic disc naturally become the main constituents of the galactic bar. Therefore, stars in the inner Galaxy that are born after the disc emergence but before the bar formation are also expected to show bar-like signatures. However, stellar populations that have ages older than the Galaxy spin-up epoch are rarely trapped by the bar and do not show bar-like signatures. We demonstrate this below using the Auriga simulations. 

\subsection{Connection between the disc formation and bar signatures}

Fig.~\ref{fig::Au1826_bar_appearing} shows the results of the analysis of the kinematics of different mono-age populations in the inner galaxy of Au18 and Au26 using the snapshot at $\tilde{z}=0$. The top panels are made with all stellar particles in Au18 and 26, while the lower set of panels focuses on particles in the inner galaxy with $|x|<5$ and $|y|<5$~kpc. The disc formation history of Au18 and Au26 are revealed using the column-normalised \citep[see e.g.][for a similar representation of the Galaxy's kinematics]{Belokurov_2020} age-circularity plot as shown in the top row of Fig.~\ref{fig::Au1826_bar_appearing}, in which the $\mathrm{circularity}=L_z/L_{\mathrm{circ}}$ is defined as the ratio of the angular-momentum in the $z$-direction to the angular-momentum of a circular orbit of equivalent energy. The orbits of the oldest stars in both galaxies are halo-like with a low and wide distribution of circularity. They have a slight prograde rotation as the mean circularity is greater than~0, similar to the results in other simulations \citep{McCluskey_2023,Chandra_2023} and in observations \citep{Chandra_2023, Zhang_2024a, Arentsen_2024}. The trend of circularity as a function of age is shown by the red-solid line. Both Au18 and Au26 have a rapid spin-up with a short timescale of $\lesssim0.5$~Gyr at a similar look-back time, $\sim11.7$~Gyr and $9.2$~Gyr ago, respectively. We denote these as the spin-up time, $t_{\mathrm{spin-up}}$, for Au18 and Au26. The circularity of stars born after the spin-up stays at a high value, marking the existence of a long-lived disc.

To study the correlation between the bar signatures and the spin-up of these two galaxies, we first compute the bar strength, $A_2$, of stars at different ages using a Fourier decomposition similar to \citet{Fragkoudi_2024} but using number density: 
\begin{align*}
    a_m(R) =  \int_0^{2\pi} \Sigma(R, \phi)\cos{(m\theta)} d\theta, \\
    b_m(R) =  \int_0^{2\pi} \Sigma(R, \phi)\sin{(m\theta)} d\theta, \\
\end{align*}
and
\begin{equation}
    A_2 = \mathrm{max}\left(\sqrt{a_2(R)^2 + b_2(R)^2}/a_0(R)\right) \text{for R<5 kpc}.
    \label{eqn::bar_strength}
\end{equation}
The bar strength of each mono-age population measured at the present day is shown by the white solid line in the same panel as the age-circularity plot in Fig.~\ref{fig::Au1826_bar_appearing}. The horizontal white dashed line that labels $A_2 = 0.25$ is the threshold chosen in \citet{Fragkoudi_2024} to mark that the bar signature becomes spatially significant. We stress that the white solid line does not represent the time evolution of the galactic bar strength. Instead, it serves as a reflection of the efficiency of bar-trapping in each mono-age population. So, the age for which the bar becomes spatially significant does not correspond to the bar formation epoch either. The bar formation time for these two galaxies as measured by \citet{Fragkoudi_2024} is marked by the vertical white dotted line. For both Au18 and Au26, we see the bar strength surpasses the threshold for the stellar population born after the spin-up because most of the kinematically cold, high-rotation stars are born after the spin-up. To further illustrate this correlation, we show the median circularity versus the bar strength for different mono-age populations in Fig.~\ref{fig::circularity_vs_A2}, in which the colour represents the age of the mono-age population. A clear positive correlation between the circularity and the bar strength of the population is evident, confirming that high-rotation stars are more easily trapped by the bar.

We closely investigate the stellar populations born after and before the spin-up by selecting stars with ages less than 1 Gyr after and before the spin-up time, $t_\mathrm{spin-up}$, and show their spatial and kinematic maps in the left and right columns of Fig.~\ref{fig::Au1826_bar_appearing}, respectively. To better compare to the Milky Way observations, we rotate the galactic bar in these two galaxies so that it is inclined at $25\degr$ to the Sun-Galactic centre line. The stellar surface density is shown in the second row of Fig.~\ref{fig::Au1826_bar_appearing}. As expected, the central surface over-density becomes more elongated for populations born after spin-up. 

However, the stellar density is hard to compare with the Milky Way observations because it is directly affected by the spatial selection function. Instead, kinematics provides a more reliable comparison. The galactic bar shows several distinct kinematic features, including i) a quadrupole pattern in the radial velocity $v_R$ field \citep{Bovy_2019, Fragkoudi_2020, Queiroz_2021}, ii) a bisymmetric pattern in the radial velocity dispersion \citep{Hey_2023, Zhang_2024} and the $|v_R|/v_{\mathrm{tot}}$ map \citep{Zhang_2024}, as well as iii) a slower angular velocity as seen in the $v_\phi/R$ map \citep{Bovy_2019}. Among these kinematic features, the quadrupole pattern in $v_R$ and the bisymmetric pattern in $|v_R|/v_\mathrm{tot}$ have the greatest contrast between the bar region and its surroundings \citep{Zhang_2024}, hence we mainly use these two maps to characterise the galactic bars in the simulated galaxies Au18 and Au26. Here, we define $v_\mathrm{tot}$ as the total in-plane velocity instead of the total 3D velocity. The radial velocity fields and $|v_R|/v_\mathrm{tot}$ fields are shown in the third and fourth rows in Fig.~\ref{fig::Au1826_bar_appearing}. We see conspicuous differences in these kinematic maps for populations born before and after the galaxy's spin-up. For both Au18 and Au26, the quadrupole radial velocity pattern and the bisymmetric $|v_R|/v_\mathrm{tot}$ pattern are seen in populations born after spin-up, but no visible bar signatures can be found for the population born before spin-up. This result supports our argument that the spin-up time is strongly correlated with and can be decoded from the kinematic signature of different mono-age populations in the inner Galaxy. We repeat the same analysis to other Auriga galaxies as well, including Au9, 10, 17, 23 and 27, which are all Milky Way-like galaxies. We reach the same conclusion for all of these galaxies except Au23, which we discuss in detail below.

\subsection{Au23 as an outlier}

The $\tilde{z}=0$ snapshot of Au23 shows a more complicated spin-up history than Au18 and Au26, as seen in the age-circularity plot in the upper left corner of Fig.~\ref{fig::Au23}. Unlike the single, quick spin-up in Au18 and Au26, we see two non-consecutive, small increases in circularity before the formation of a long-lived disc. The ages of the two small increments in circularity are denoted by the vertical red-dashed and red-dotted lines, and we dub them $t_1$ for the increment at the younger age and $t_2$ for the one at the older age ($t_1 \sim 9.6$~Gyr and $t_2 \sim 11.6$~Gyr). The evolution of the bar strength reveals that the bar signatures appear around $t_2$ instead of $t_1$, which is long before the disc-formation time as seen at $\tilde{z}=0$. The kinematics of stars reveal the same results. We select stars born within 1 Gyr after $t_1$, between $t_1$ and $t_2$, and 1 Gyr before $t_2$ and present the same spatial and kinematics maps as Fig.~\ref{fig::Au1826_bar_appearing} for the left, middle, and right panels, respectively. The bar signatures can be seen in the left and middle columns, but the population born before $t_2$ does not show bar-like kinematics. One interpretation of this result is that abundant stars born between $t_1$ and $t_2$ are trapped by the galactic bar despite their hot kinematics and slow rotation. However, a different picture is seen in an earlier snapshot of Au23 as we discuss below.

The bar formation in Au23 occurred $5.06$~Gyr ago according to \citet{Fragkoudi_2024}. We inspect the snapshot at $\tilde{z}=0.505$ ($t_{\mathrm{lookback}}\sim5.2$~Gyr), which is one of the closest snapshots before the bar formation time of Au23. The age-circularity trend in that snapshot is shown in the upper right corner of Fig.~\ref{fig::Au23}. From this snapshot, we see that the spin-up of Au23 occurs at $t_2$ instead of $t_1$. At the time of bar formation, a higher fraction of stars born between $t_1$ and $t_2$ has cold, disc-like kinematics compared to that fraction seen at $\tilde{z}=0$. The population born between $t_1$ and $t_2$ are heated into halo-like stars by merger events at $t_{\mathrm{lookback}}\sim3$~Gyr in Au23. We do not further discuss the properties of these merger events, as it is not the focus of this paper. In the disc region, these old disc stars are disrupted into halo-like orbits, but for those in the inner galaxy, their bar-like kinematic signatures persisted even after the merger. Hence, Au23, as an example, illustrates the possibility of probing an old, highly-damaged or disrupted disc in the Milky Way if that exists. 

With the aid of the snapshot at $\tilde{z}=0.504$ in Au23, we find that Au23 is not in contradiction with the hypothesis that the maximum age of populations exhibiting bar signatures can still be used to measure the disc formation time in a galaxy. However, Au23 shows that this measurement should be considered as a constraint on the upper age bound of the galactic disc we see today.

\subsection{Bar signature in the $v_\ell$ map}
\label{sec::vl_in_simulation}

\begin{figure}
    \centering
    \includegraphics[width=\columnwidth]{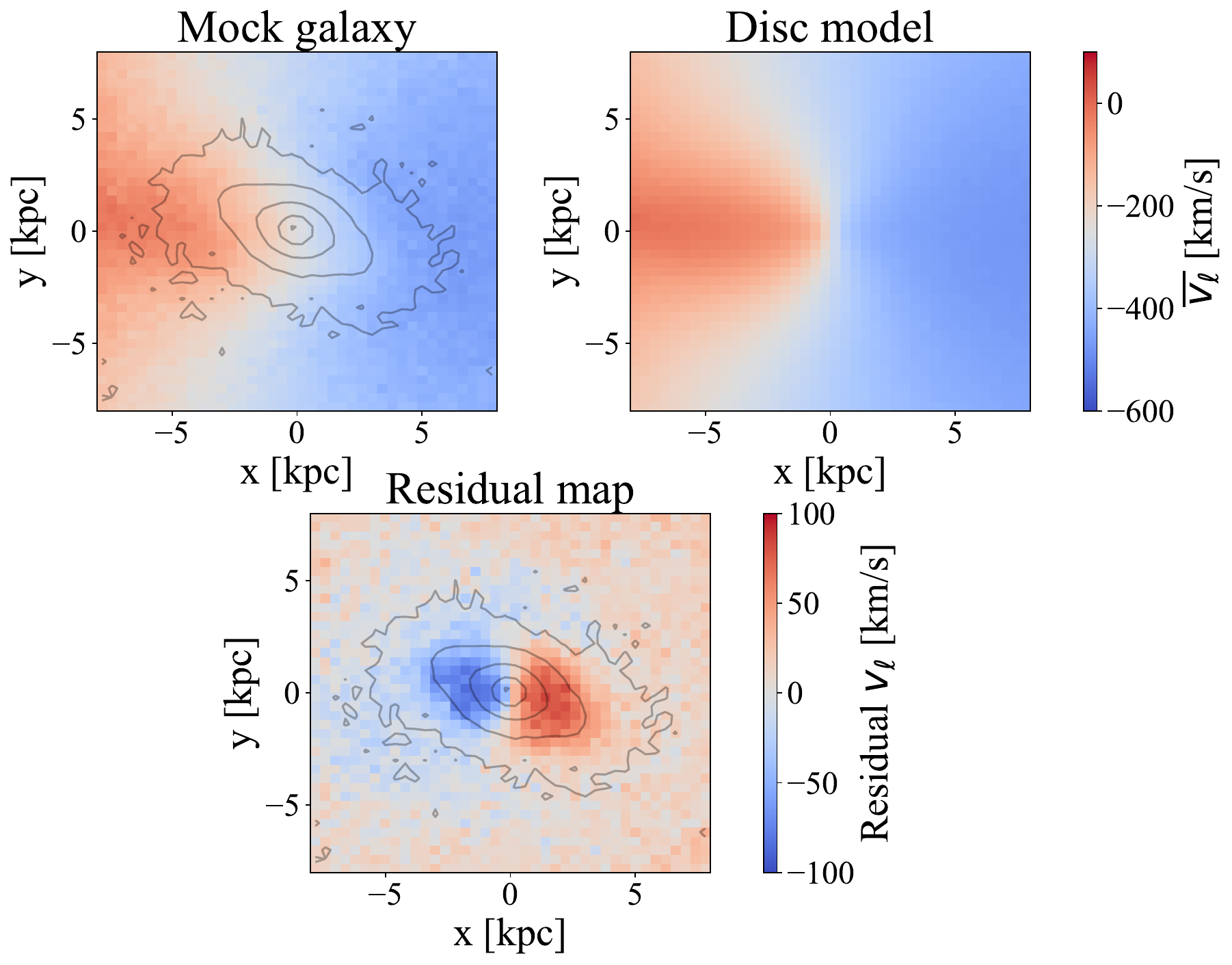}
    \caption{An illustration on the effects of the galactic bar on the $v_\ell$ field of an N-body simulated galaxy. \textit{Top left:} the $v_\ell$ field of the mock galaxy. \textit{Top right:} the $v_\ell$ field of the corresponding disc model we constructed for this mock galaxy. \textit{Bottom:} the residual, $\Delta v_\ell$, map, in which we find an inclined, dipole-like signature that corresponds to the galactic bar. The black contours show the surface density distribution of stellar particles in this galaxy.}
    \label{fig::vl_Nbody}
\end{figure}

\begin{figure*}
    \centering
    \includegraphics[width=\textwidth]{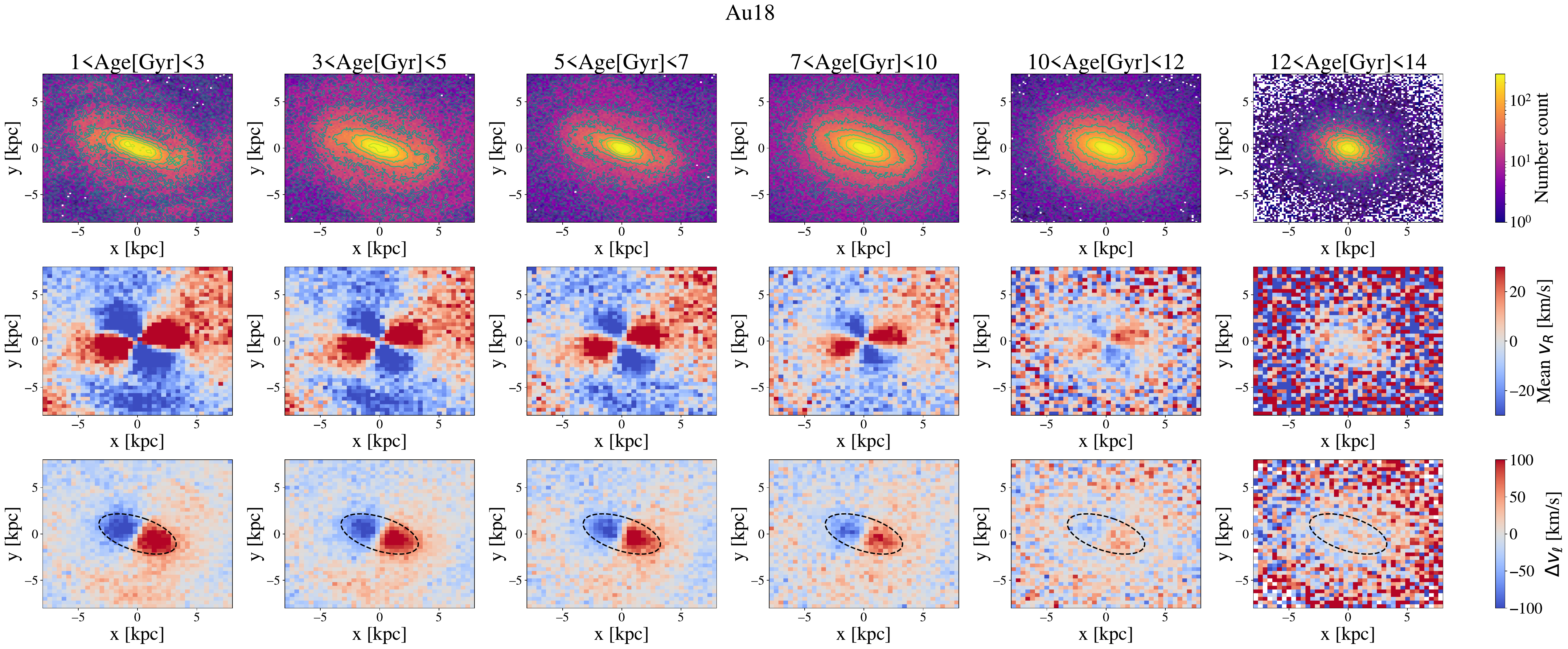}
    \caption{A demonstration of using the $v_\ell$ map to trace the chrono-kinematic evolution of the inner galaxy of Au18. Each column shows the spatial and kinematic fields of the stellar population at different ages. \textit{Top row:} the surface density distribution. \textit{Middle row:} the mean radial velocity fields of stellar populations in Au18 with different ages. \textit{Bottom row:} the residual $v_\ell$ map, for which a different disc model is fitted for each age bin. The black dashed ellipse is used to illustrate the angle of the galactic bar in Au18 in this snapshot. We highlight that the trend of the strength of the dipole signal is similar to that of the quadrupole radial velocity pattern, and the residual dipole pattern is gone when there is no quadrupole pattern in the radial velocity map.}
    \label{fig::vl_Au18}
\end{figure*}

Because the Miras' line-of-sight velocity measurements are missing from \textit{Gaia} DR3 due to the variables' high pulsation amplitude, we cannot access the full 6D phase space of these stars, at least presently. Since the radial velocity map and the $|v_R|/v_\mathrm{tot}$ map are not available for our Mira sample (described in Section~\ref{sec::data}), we hunt for possible bar signatures hiding in the $v_\ell (= \mu_l\times d)$ map instead. Note that $v_\ell$ is heliocentric, and not the Galactocentric longitudinal velocity component of the star because we do not correct it for the Solar motion. 

We use an N-body simulation snapshot to demonstrate the behaviour of the galactic bar in the $v_\ell$ map \citep[Galaxy B in][]{Zhang_2024}. The N-body simulation is set up using the initial conditions in \citet{TG21}, and include a disc, a stellar bulge and a dark matter halo with masses and sizes similar to those of the Milky Way \citep{BH_2016}. The snapshot is taken at 4 Gyr after the beginning of the simulation, at which time a stable and long-lived bar is present. Assuming a Solar reflex motion of ($12.9$, $245.6$, $7.78$)~km/s in the radial, azimuthal, and vertical directions, respectively \citep{Schonrich_2010}, we transform to galactic coordinates and map the $v_\ell$ field of the simulated galaxy as shown in the top left panel in Fig.~\ref{fig::vl_Nbody}. The overall trends in the $v_\ell$ map are due to the galactic rotation. Hence, to isolate bar signatures, we subtract the $v_\ell$ caused by the rotation of the disc. To subtract the background rotation for this N-body simulated galaxy, we use the quasi-isothermal disc model \citep{Binney_2010} with the same parameters as we used to generate the initial condition. We generate stars from the quasi-isothermal disc distribution function and use this to map the $v_\ell$ distribution of the disc. The $v_\ell$ map of the disc model is shown in the top right panel of Fig.~\ref{fig::vl_Nbody}. Subtracting the mean $v_\ell$ of the disc model from the mock galaxy for each $x-y$ pixel, the residual $\Delta v_\ell$ is shown in the bottom panel. A dipole-like signature with a similar orientation and size of the galactic bar appears because the galactic bar is rotating slowly compared to the disc (also see Fig.~2 and Fig.~3 in \citealt{Bovy_2019}).

To further verify the applicability of using the $v_\ell$ map to explore the kinematic signature of the galactic bar with different mono-age populations, we perform a test using Au18 by binning stars into age segments. The surface density and radial velocity maps are shown in the first and second rows of Fig.~\ref{fig::vl_Au18}. Recalling that $t_\mathrm{spin-up} \sim 11.7$ Gyr, no bar signatures are expected for the stellar population with $12<\mathrm{Age/Gyr}<14$. A bar signature appears for the population younger than 12 Gyr and becomes stronger at younger ages. To calculate the $\Delta v_\ell$ map, a disc model is needed for each age bin. We fit a quasi-isothermal disc for stars between $5<R/\mathrm{kpc}<12$ and $|z|<3$ kpc by maximising the log-likelihood:
\begin{equation}
    \ln L = \sum_{i=1}^N \ln p_i(\mathbf{v}|\mathbf{x}),
    \label{eqn::logL_easyform}
\end{equation}
where
\begin{equation}
    p(\mathbf{v}|\mathbf{x}) = \frac{f(\mathbf{J})}{\int \mathrm{d}^3\mathbf{v} f(\mathbf{J})}.
\end{equation}
$f(\mathbf{J})$ is the quasi-isothermal distribution function \citep{Binney_2010}, where $\mathbf{J}(\mathbf{x,v})$ is the action \citep{Sanders_2016}. The potential of the galaxy is computed using the \texttt{CylSpline} approximation method in \textsc{Agama} \citep{Vasiliev_2019}. Subtracting the $v_\ell$ maps of the fitted quasi-isothermal discs from the original $v_\ell$ maps in different age segments, we present the residuals in the bottom row in Fig.~\ref{fig::vl_Au18}. Similar to the stellar density and radial velocity map, a strong bar (dipole) signature is seen in the young population and becomes weaker for the older population. No visible dipole signal can be found in the bar region for the population older than $12$~Gyr, and the dipole strength there is consistent with zero. This test validates the use of the dipole pattern in the $v_\ell$ residual map to study the chrono-kinematics of the inner Galaxy.

\section{Spin-up epoch of the Milky Way}
\label{sec::spin_up_epoch_MW}

\begin{figure*}
    \centering
    \includegraphics[width=\textwidth]{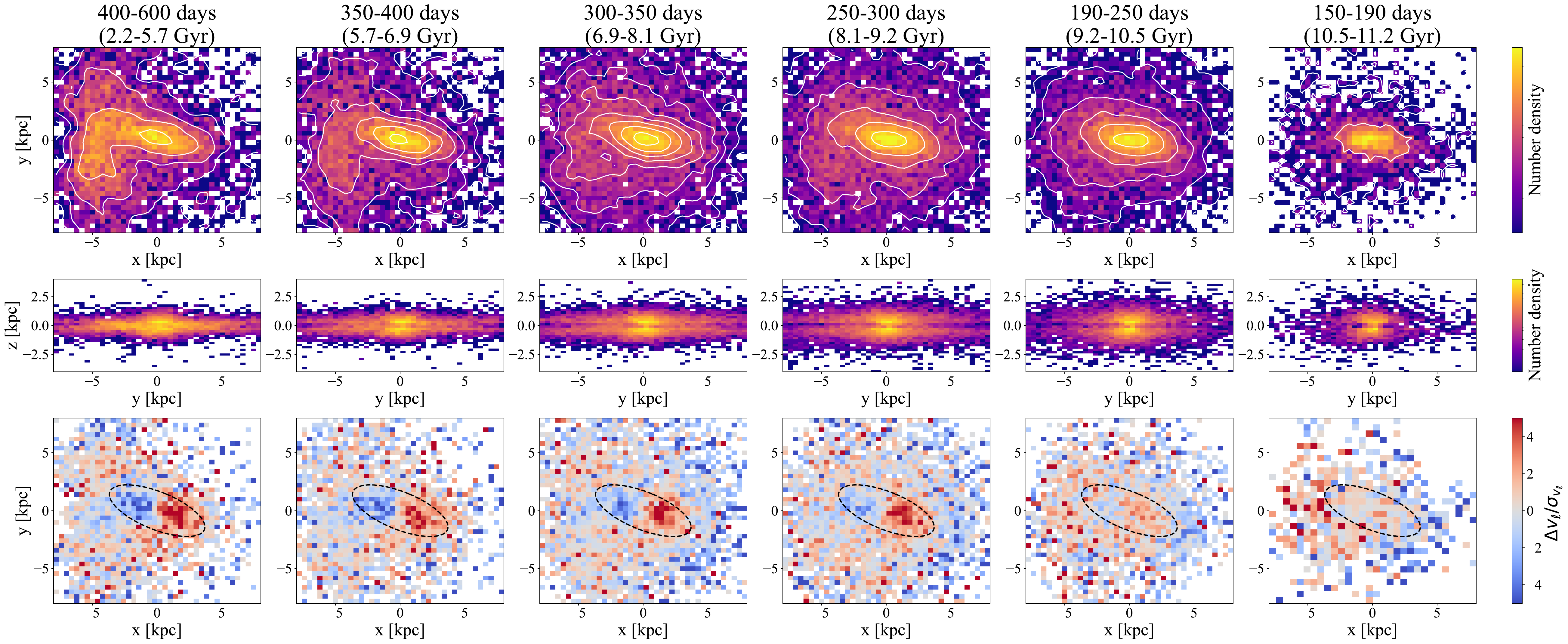}
    \caption{The spatial and kinematic signatures of O-rich Mira variables candidates in different period segments. \textit{Top row:} the surface density distribution. \textit{Middle row:} the side-on density projection in the $y-z$ plane. \textit{Bottom row:} the residual $v_\ell$ map but divided by the Poisson uncertainty. The disc model used for each period bin is from the fitted quasi-isothermal disc in \citet{ZS23}. The black dashed ellipse shows the hypothetical inclination and the size of the Milky Way bar, which is in good agreement with the dipole signal in the residual $v_\ell$ map when that exists. }
    \label{fig::Mira_bin}
\end{figure*}

We split our Mira variable sample into different period bins, which is analogous to binning them by age thanks to the period-age relation of Mira variables \citep{Wyatt_Cahn_1983,Feast_2009,Grady_2019,ZS23}. The behaviour of Mira candidates in configuration space is shown in the first and second rows in Fig.~\ref{fig::Mira_bin}. From the spatial distribution in the $(x,y)$ plane, an inclined, bar-like structure is seen in the period bins with $\mathrm{period}>190$ days. A bar-like overdensity is also found in populations with a period of 150-190 days but instead it clearly aligns with the $x$-axis, similar to the results in \citet{Grady_2020}. Distance uncertainty can bias the inclination angle of the Galactic bar by smearing the bar stars along the line-of-sight direction, and hence, shearing the bar so that it becomes more aligned with the Sun-Galactic centre (GC) line \citep{Hey_2023, Zhang_2024}. However, to make the bar completely aligned with the Sun-GC line, distance uncertainty should be $25\%-30\%$, whereas for our sample, the distance uncertainty is only $\sim10\%-15\%$ as shown in Fig.~\ref{fig::disterr_xy_heating}. 
A distance uncertainty on the order of $10\%$ can also \emph{create} an elongated, bar-like overdensity that is aligned with the Sun-GC line from an approximately spherical distribution. We demonstrate this in Appendix~\ref{Appendix::distance_uncertainty} with details.
In brief, as the heliocentric distance uncertainty displaces and smears stars along the line-of-sight direction, it stretches a spherical-like overdensity into a bar-like overdensity oriented along the line-of-sight. Therefore, the horizontal (in $x-y$ projection) bar-like signature we find in the Mira variables with a period of 150-190 days could be associated with a quasi-spherical overdensity in the inner Galaxy, for example, a spherical bulge or inner halo.

From the side-on density projection shown in the second row of Fig.~\ref{fig::Mira_bin}, a larger scale height is found for Mira candidates with shorter periods (corresponding to the older population), which agrees with the results in \citet{Queiroz_2023}. Directly discussing the existence of the disc components from the spatial configuration of these populations is challenging as the age uncertainty affects the conclusion, as discussed in detail in Section~\ref{sec::age_uncertainty}. Hence, we avoid over-interpreting the side-on projections of the Mira variables with a period between 150-190 days. 

To examine the signature of the Galactic bar in different mono-age stellar populations in the inner Milky Way with the $v_\ell$ map, we stick to the quasi-isothermal disc distribution function as the basis of the disc model in each period bin. We adopt the fitted distribution function parameters from Table 1 in \citet{ZS23}. \citet{ZS23} also selected Mira candidates from the \textit{Gaia} LPV catalogue with very similar filters and cuts, except only stars with $5<R/\mathrm{kpc}<10$ are considered in that work (as the focus was the local age--velocity dispersion relation). Hence, the Mira variable sample we have constructed in this work in the region $5<R/\mathrm{kpc}<10$ should have the same kinematic properties as those in \citet{ZS23}. In \citet{ZS23}, we split the Mira sample into period bins and fit a quasi-isothermal distribution function for stars in each bin. The fitting is performed by maximising a log-likelihood similar to Eq.~\ref{eqn::logL_easyform}, but the measurement uncertainties and the missing line-of-sight velocities are also taken into account by marginalisation.

The width of the period bin in \citet{ZS23} is generally smaller than the period bin adopted in this work, so the difference in the binning strategy has to be considered when comparing the disc model to the observation. For each of our period bins (which covers 2-3 period bins in \citealt{ZS23}) mock stars are generated from several disc models, for which the fractional contribution from each disc model is assigned according to the period distribution in our period bin. 
The mock stars are generated with the fitted quasi-isothermal distribution function parameters in \citet{ZS23} using the \textsc{Agama} \texttt{Sampling} routine (see details in \citealt{Vasiliev_2019}).
To fairly compare the $v_\ell$ map of the disc model and the observation, we scatter mock generated disc stars with $12\%$ distance uncertainty. Then, we match the distribution of the mock stars to the observed spatial distribution of our sample by finding the closest matched mock star for each observed star in the configuration space. This routine is repeated five times to minimise the effect of Poisson uncertainty on the disc model. We then subtract the $v_\ell$ map of the disc model from the observation. In the lower row of Fig.~\ref{fig::Mira_bin}, we show the residual $v_\ell$ map in terms of $\Delta v_\ell/\sigma_{v_\ell}$ for each period bin, where $\sigma_{v_\ell}$ is the Poisson error for each $(x,y)$ pixel. A strong dipole signal that is inclined like the Galactic bar is seen in the long period (young) population as expected due to the Galactic bar. By visual inspection, the dipole signal gradually becomes weaker for a shorter period (older) population and no dipole-like signature is noticeable in the 150-190 period Mira variables. 

\begin{figure}
    \centering
    \includegraphics[width=\columnwidth]{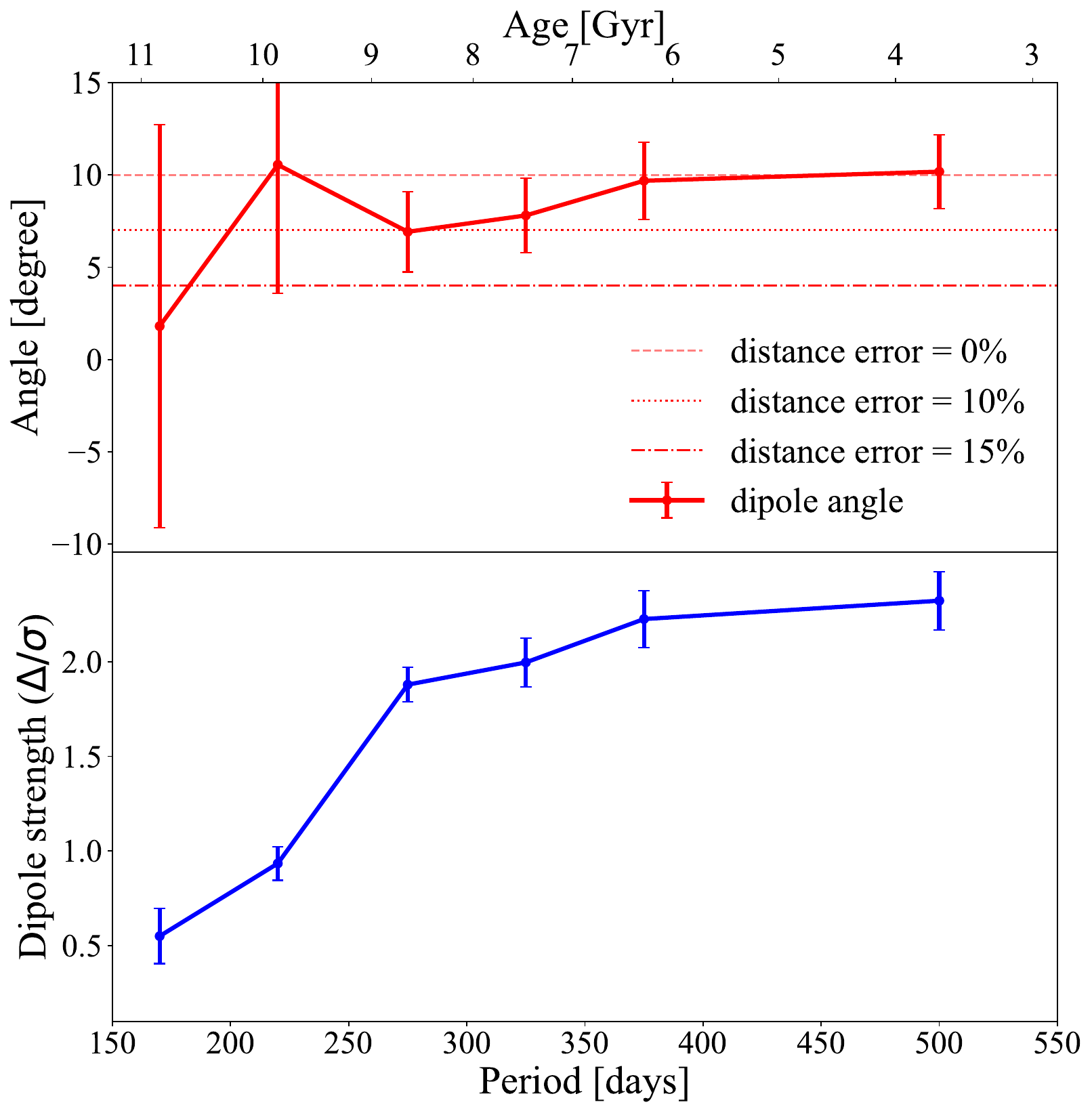}
    \caption{
    \textit{Top:} The dipole angle for the residual $v_\ell$ map for O-rich Mira variables in different period bins. The dashed, dotted, dash-dotted red lines in the top panel are the expected dipole inclination in the residual $v_\ell$ map if the signal is contributed by the galactic bar. The uncertainty in the top panel is composed of statistical and observational error, in which the statistical uncertainty comes from the standard deviation in the angles determined from all radial annuli. \textit{Bottom:} The dipole angle for the residual $v_\ell$ map for O-rich Mira variables in different period bins. The dipole strength is normalised by the Poisson uncertainty. The uncertainty for dipole strength is from observational error that is propagated by 100 Monte Carlo realisations of the observations.}
    \label{fig::angle_strength}
\end{figure}

To quantify the dipole signal, we compute the Fourier components for the $\Delta v_\ell/\sigma_{v_\ell}$ map as 
\begin{align*}
    a_1(R) = \int_0^{2\pi} \frac{\Delta v_\ell}{\sigma_{v_\ell}} \cos{(\theta)} \mathrm{d}\theta, \\
    b_1(R) = \int_0^{2\pi} \frac{\Delta v_\ell}{\sigma_{v_\ell}}  \sin{(\theta)} \mathrm{d}\theta.
\end{align*}
Here, the dipole angle is $0.5\tan^{-1}(b_1/a_1)$ and the strength is $\sqrt(a_1^2 + b_1^2)$ for each radial annulus. We pick the width of the radial annulus to be 0.4 kpc. For each period bin, we take the average of the dipole angle for each annulus inside 4 kpc, and show the average angle associated with each period bin in the top panel of Fig.~\ref{fig::angle_strength}. The final dipole strength shown in the lower panel is calculated by taking the maximum dipole strength of annuli inside 4 kpc. The observational uncertainties are propagated to the angle and the strength with 100 Monte Carlo realisations. The uncertainty associated with the angle is contributed by both statistical and measurement uncertainty, while the uncertainty of the dipole strength is purely from observational errors. The red, horizontal, dashed lines in the top panels of Fig.~\ref{fig::angle_strength} are the inclination angle of the dipole signature in the N-body galaxy in Section~\ref{sec::vl_in_simulation} under different distance uncertainty, and we use those as aids to interpret the dipole angle. We confirm that the dipole patterns for all period bins greater than 190 days are indeed inclined and consistent with the expected dipole angle induced by the Galactic bar. The dipole angle for Miras with 150-190 periods shows an angle close to zero accompanied with a large uncertainty. This implies that no consistent dipole angle can be found across all annuli, so no clear bar signature exists in that period bin. The $\Delta v_\ell$ dipole signal quickly grows for periods greater than 250 days. Therefore, quantifying the dipole signature with Fourier decomposition, the results are still consistent with the conclusion that no bar signal is found for Miras with 150-190 days period using the $v_\ell$ map. 

According to the analysis with simulated galaxies in Section~\ref{sec::chrono_kinematic_simulation}, we can associate the oldest population that exhibit bar signatures with the spin-up time of the Milky Way. Hence, the Mira variables with a period smaller than 190 days are likely born before the Milky Way spin-up epoch. Using the period-age relation of Mira variables, we can constrain the spin-up time of the Milky Way. The period-age relation presented in \citet{ZS23} has the functional form:
\begin{equation}
\tau = \frac{\tau_0}{2}\left(1+ \tanh\Big[\frac{330-P(\mathrm{days})}{P_s}\Big] \right),
\label{eqn::period-age-relationship}
\end{equation}
where $\tau_0$ and $P_s$ are the associated parameters. \citet{ZS23} fitted Eq.~\ref{eqn::period-age-relationship} and found $(\tau_0, P_s)$ = (13.7 Gyr, 401 days) when considering Mira kinematics only or $(\tau_0, P_s)$ = (14.7 Gyr, 308 days) when information of Mira variables in globular clusters is included. \citet{ZS23} found an age scatter of $11\%$ at a fixed period based on the difference in the Mira ages based on kinematics and on globular cluster ages. In this work, we adopt the $11\%$ scatter as the intrinsic dispersion in the period-age relation. However, the true intrinsic dispersion may require further calibration as theoretical models predict a significant intrinsic scatter \citep{Trabucchi_2019}. Mira variables with a period of 190 days have ages corresponding to $10.5\pm1.2$~Gyr and $9.1\pm1.0$~Gyr for $(\tau_0, P_s)$ = (14.7 Gyr, 308 days) and $(\tau_0, P_s)$ = (13.7 Gyr, 401 days), respectively. Therefore, the age of the Milky Way disc is likely to be younger than $\sim12$~Gyr (10.5 + 1.2 Gyr). This result is also consistent with the observed age distribution of the Galactic bulge in \citet{Bensby2013}.

\section{Discussion}
\label{sec::discussion}

We have analysed kinematic properties of the O-rich Mira variables with different periods and have used these to constrain the Milky Way spin-up epoch in Section~\ref{sec::spin_up_epoch_MW}. We now compare our method and the results to other lines of evidence of the spin-up time of the Milky Way, those obtained either with metallicity or directly through stellar ages. We also discuss the influence of age uncertainty on the performance of our method. Additionally, we address possible caveats of the approach employed in this work, i.e.  using $v_\ell$ maps as a probe of the bar signal.

\subsection{Comparison with other estimates of the disc formation time}

\subsubsection{The very metal-poor (VMP) disc}

\citet{Sestito_2019, Sestito_2020} found that a large fraction of VMP stars with metallicity $\mathrm{[M/H]}\lesssim-2.5$ have disc-like kinematics. Several possible scenarios for the existence of VMP stars with disc-like orbits are discussed in \citet{Sestito_2020}. Of these, albeit the least likely and clearly disfavoured by the authors, is that in which the Milky Way disc was already in place during the very metal-poor era. \citet{Di_Matteo_2020} argued that stars with disc-like kinematics can be found ubiquitously across the entire metallicity range reaching $\mathrm{[M/H]}\sim-6$, thus favouring a scenario of disc formation at very low metallicities. 
Note that, according to numerical simulations, the MW managed to form a coherently rotating, dominant stellar disc one of the earliest amongst the galaxies of similar mass \citep[see e.g.][]{Semenov_2024, Dillamore_2024}. However, given the mass and metallicity evolution of the Galaxy, its ancient disc stars were not metal-poor.

\citet{Belokurov_2022} used the APOGEE survey data to show that the azimuthal velocity of \textit{in-situ} stars increases sharply around $\mathrm{[M/H]}\sim-1.3$. \citet{Chandra_2023} and \citet{Zhang_2024a} found similar results using \textit{Gaia} XP metallicities. \citet{Zhang_2024a} demonstrated that VMP stars on disc-like orbits could simply be a part of the prograde halo, and that a non-negligible fraction of the halo stars in the Solar neighbourhood could have disc-like orbital parameters without being part of a distinct disc "component". \citet{Dillamore2023,Yuan_2023, Li_2023} showed that a slowing-down Galactic bar could drag halo stars from the inner Milky Way closer to the Solar radius, depositing them on disc-like orbits. Based on the age-metallicity relation of the Milky Way \citep{Haywood_2013, Xiang_2022}, a Galactic disc that formed around $\mathrm{[M/H]}\sim-1$ to $-1.5$ ($11.5\lesssim\mathrm{age/Gyr}\lesssim12.5$) is consistent with our constraint on the disc formation time using Mira kinematics of the inner Galaxy. 

\begin{figure*}
    \centering
    \includegraphics[width = 0.49\textwidth]{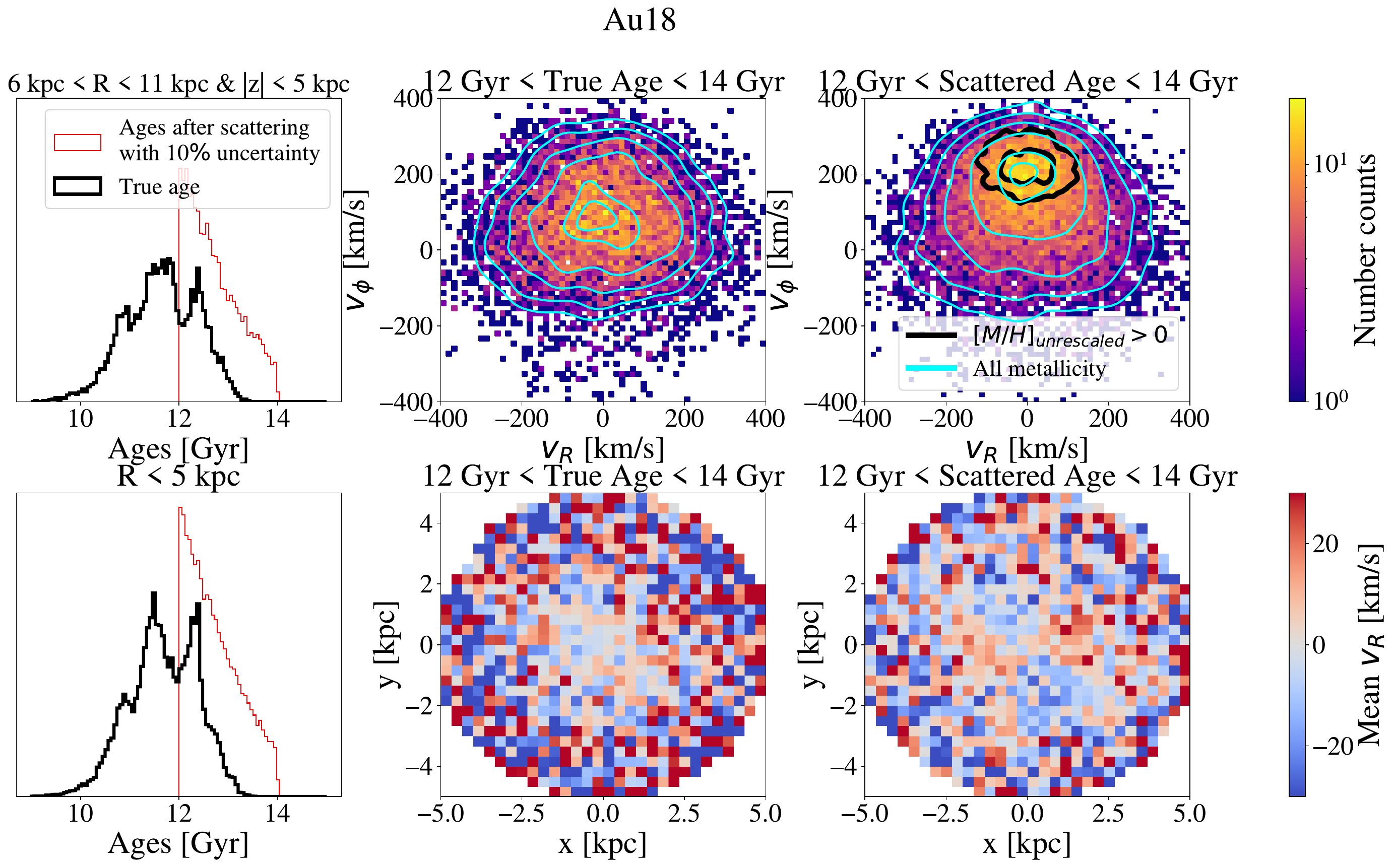}
    \includegraphics[width = 0.49\textwidth]{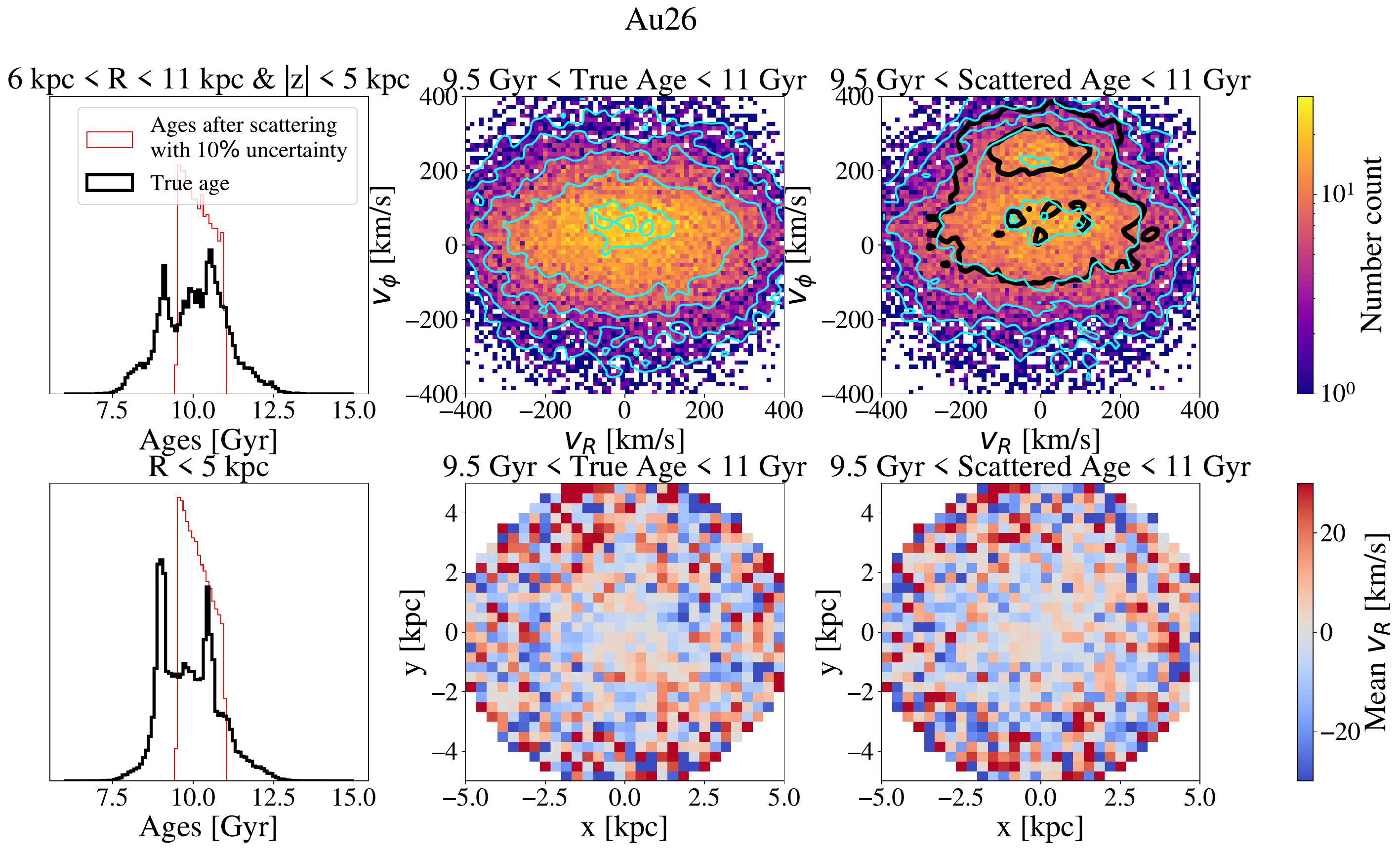}
    \caption{The effects of the age uncertainty in determining the disc formation time. The left sets of panels are for Au18, and right sets are for Au26. In each sets of figures, the \textit{top row} focuses on stars on the disc, which $5<R<12$ and $|z|<5$ kpc, and the $v_R$ vs. $v_\phi$ plane is used to distinguish disc and halo population; the \textit{bottom row} focuses on inner galaxy stars with $R<5$~kpc, and the radial velocity map is used to show the disc star appearance. \textit{Left column:} the age histograms for the old stars selected from uncertainty scattered age. The distribution of the scattered age is in red, and the true age is in black. The red histogram is truncated because we only focus on the stars we binned with the uncertainty-added age, so the truncation of the red histogram corresponds to the age bin. \textit{Middle column:} the kinematics of the true old stars born before the spin-up. \textit{Right column:} the kinematics of the old stars selected with age after scattering with $10\%$ uncertainty. The aqua contours are for highlighting the distribution of all stars in the $v_R$ vs. $v_\phi$ space, and the black contours are used to illustrate the distribution of metal-rich stars.}
    \label{fig::age_uncertainty}
\end{figure*}

\subsubsection{Age of the disc}

The age distribution of disc stars provides information on the age of the Galactic disc. Ages of individual stars can be measured using isochrone fitting \citep{Sanders_Das_2018, Queiroz_2018, Xiang_2022, Nataf_2024}, data-driven neural networks \citep{Mackereth_2019}, and asteroseismology \citep{Miglio_2021, Warfield_2024} with different precision and accuracy, respectively.

\citet{Queiroz_2023} applied the t-distributed stochastic neighbour embedding (t-SNE) to the chemical abundances measured by APOGEE and used a clustering algorithm to select a population chemically similar to the thick disc. The isochrone ages of the thick disc stars are generally old (their distribution peaks at $\mathrm{age}\gtrsim10.9$~Gyr). Although their results are not in direct contradiction with our constraints on the disc formation time, we discuss and demonstrate in Section~\ref{sec::age_uncertainty} below that age uncertainty can produce an overestimation of the disc age measurements.
Similarly, \citet{Miglio_2021} found an age of $\sim11$~Gyr after deconvolving the age uncertainty for the chemically-defined thick disc populations with ages from asteroseismological measurements using Kepler data. \citet{Gallart_2024} designed a colour-magnitude diagram(CMD)-fitting technique to reveal the star formation history of thin disc stars in the solar neighbourhood ($d\lesssim100$ pc). They found that the star formation of the disc started around $\sim11-10.5$~Gyr ago. \citet{Belokurov_2024} separated the globular clusters into \textit{in-situ} and accreted with an energy-angular momentum cut. The \textit{in-situ} globular clusters spun up, i.e. increased dramatically their bulk azimuthal velocity at $-1.3\lesssim\mathrm{[Fe/H]}\lesssim-1$ around $11.7-12.7$~Gyr ago, which is also in broad agreement with the result in this work.

\citet{Nepal_2024} used the \texttt{StarHorse} Bayesian isochrone ages \citep{Queiroz_2018} with stellar parameters from the hybrid-CNN analysis of the \textit{Gaia} RVS sample \citep{Guiglion_2024} and found a kinematically cold, thin disc-like population emerging within 1 Gyr after the Big Bang. Their old, thin disc-like population covers a large range in metallicity from metal-poor to super-metal-rich and is particularly conspicuous at super-solar metallicity. It is difficult to reach super-metal-rich metallicities in the Galactic disc region within 1 Gyr after the Big Bang. It could be therefore that some (or many) of these old disc stars have come from the inner Galaxy, and thus are related to effects of the bar \citep{Dillamore2023,Yuan_2023, Li_2023, Nepal2024a, Dillamore_2024b}. 

We emphasise that our measurements are not influenced by the bar formation time in the Milky Way. In all the Auriga galaxies we inspected, we found that variation in time difference between bar formation and disc formation does not affect our central argument that the oldest stellar population showing bar signatures is the first population born after disc formation. As mentioned in Section~\ref{sec::chrono_kinematic_simulation}, the disc population born before bar formation becomes trapped by the Galactic bar, regardless of when the bar formed. Therefore, bar chrono-kinematics are not suitable for distinguishing the bar formation time, whether it happened around $\sim$3~Gyr ago, as suggested by \citet{Nepal2024a}, or 7-10~Gyr ago, as suggested by \citet{Sanders_2024, Haywood2024}.


\subsection{Effects from the age uncertainty}
\label{sec::age_uncertainty}

Scatter due to the age measurement uncertainty causes younger stars to be confused for apparently old stars. As the early Galaxy built its stellar mass up and the star formation rate increased, significantly more stars formed during and after the spin-up of the Milky Way than before. Therefore, the population born before the Galaxy spin-up are a relative minority compared to the subsequent stellar generations. These younger stars can produce noticeable contamination in a given bin of measured old age if the age uncertainty is non-negligible.
We demonstrate the effects of such contamination using model galaxies Au18 and Au26 introduced earlier, as shown in Fig.~\ref{fig::age_uncertainty}. 

For each simulated galaxy, we add $10\%$ scatter to the ages of stars to reproduce best currently available observations. Recalling that Au18 and Au26 spin up rapidly around $11.7$ and $9.2$~Gyr ago, we select stars with observed, i.e. scattered, ages between $12-14$~Gyr for Au18 and $9.5-11$~Gyr for Au26, respectively. In the top rows of the Figure, we focus on the disc region in these galaxies, $6<R<11$~kpc and $|z|<5$~kpc, while the lower rows show stars in the inner galaxy ($R<5$~kpc). The 1D distributions of the scattered age and the true age of the selected stars are shown as the red and black histogram in the left columns. For both the disc region and the inner galaxy region, younger star contamination of $\sim30-70\%$ is seen. The fraction of the contamination depends on the detailed star-formation history of each galaxy. 

To illustrate how this contamination can cause overestimation of the disc age, we bin stars with their true ages and the uncertainty scattered ages in an age bin before the spin-up. In the disc region, we use the $v_R$ vs. $v_\phi$ map to distinguish the halo and the disc population. Focusing on the true age population we see that before the galaxy's spin-up, no rotation-supported component is found, as expected. However, using the same age bin but now applied to stars with scattered ages, a fast-rotating, disc-like population appears, which can be seen from the aqua contours of the number distribution in the $v_R$ vs. $v_\phi$ map. This apparently old disc-like component is solely due to the contamination by the younger stars. The disc population is more conspicuous when we further select metal-rich stars, as shown by the black contour. Therefore, binning stars by age in the presence of age uncertainty, results in the disc population appearing noticeably older than it actually is.


The same contamination, and hence the caveat, also exists in our study but is less pronounced as we demonstrate below. For Au18, because of the contamination, the quadrupole radial velocity pattern also appears if we select old stars after adding the age scatter  (bottom right panel in the Au18 panels on the left-hand side of Figure~\ref{fig::age_uncertainty}). The contamination fraction here is $\sim65\%$ for Au18 in the inner galaxy. However, in Au26, no bar signature is seen even when we select the age bin before spin-up. The younger disc star contamination fraction for Au26 in the inner Galaxy is $\sim35\%$. At the same contamination fraction, a fast rotating disc population can be identified in the disc region of Au26, as we see in the corresponding panel in the upper row. Therefore, we argue that our method is comparably less sensitive to the younger disc star contamination. This is because the kinematics of stars with disc- and halo-origins are more similar and less distinguishable from each other in the inner Galaxy compared to the disc region. 

Also, if we treat the spin-up time obtained in this work as an upper limit, then this caveat does not affect the conclusion that the Milky Way disc is younger than 12 Gyr. Forward modelling the data using both age and distance uncertainties should provide a more accurate and robust way to determine the true spin-up epoch, and we leave this for future work. 

\subsection{Caveat of using $v_\ell$ map}

Due to the unavailability of the line-of-sight velocity measurements  of the Mira variables, we do not have the full 6D phase space information for the inner Galaxy. Accordingly, our search for potential bar signatures is limited to the $v_\ell$ map. As mentioned in Section~\ref{sec::vl_in_simulation}, the trends in the $v_\ell$ maps are mainly due to the galactic rotation. The dipole signal found for the bar in Fig.~\ref{fig::vl_Nbody}, \ref{fig::vl_Au18}, and \ref{fig::Mira_bin} is due to bar rotating slower than the disc in the inner galaxy. Therefore, the $v_\ell$ map loses the sensitivity to the bar kinematics when in the inner Galaxy the disc rotation becomes comparable to that of the bar. However, inspecting the Auriga galaxies, we find no evidence that the bar signatures can exist in a mono-age population where the rotation is comparable to or slower than the rotation of the bar. Hence, this caveat is likely unimportant. Using kinematic maps that trace the special kinematic signature of bar-supporting orbits, such as the radial velocity or $\left|v_R/v_\mathrm{tot}\right|$ map, would be preferred when full 6D phase space information is available. 

In the upcoming \textit{Gaia} DR4 and 4MOST survey \citep{4MOST}, this problem could be resolved with a sample of precise age and distance measurements in the inner Galaxy. Alternatively, with the radial velocity time-series for all LPV candidates in \textit{Gaia} DR4, we can access the line-of-sight velocities for Mira variables and correct them for pulsation. With a complete set of 6D phase-space coordinates in hand, a model of the chrono-kinematic signatures of the Galactic bar with the radial velocity or $v_R/v_\mathrm{tot}$ map can be built, which would be more robust compared to the $v_\ell$ map. 

\section{Conclusions}
\label{sec::conclusion}

In this work, we have presented a novel method to constrain the spin-up epoch of the Milky Way using kinematics of stars in the inner Galaxy. Using Auriga simulations, we demonstrate that the disc formation time of a barred galaxy can be inferred from the presence of bar-like kinematics in mono-age populations. We use Au18 and Au26 as example model galaxies, which both experienced a rapid spin-up, similar to what is observed in the Milky Way \citep{Belokurov_2022,Chandra_2023, Zhang_2024a}. In these simulated objects, both spatial and kinematic signatures of the bar are detectable in the stellar populations born after the spin-up. This is because kinematically cold stars respond more readily to the bar compared to the kinematically hot stars. These kinematically cold stars, born soon after the rapid disc formation, display bar-like kinematics, indicating that the ages of the oldest stellar population with bar signatures match the disc spin-up epoch rather than the bar-formation epoch.
Note that while we only discuss Au18 and Au26 in this work, the same results are also found in other Milky Way-analogue Auriga galaxies.

We construct an O-rich Mira variable sample from the \textit{Gaia} LPV candidates \citep{GaiaDR3_LPV}. O-rich Mira variables are useful age indicators thanks to their period-age relation \citep{Grady_2019, ZS23}. Due to the missing line-of-sight velocity measurements for Mira variables, we use the $v_\ell$ map to probe the signal from the Galactic bar. To emphasise the bar signatures in the $v_\ell$ map, we remove the systematic trends in $v_\ell$ map caused by the Galactic rotation and focus on the residual. Using a model galaxy created with an N-body simulation, we show that a dipole-like signal with an inclination and a shape similar to those of the galactic bar can be identified in the residual, $\Delta v_\ell$, map. This is because in the inner Galaxy, the galactic bar is rotating slower than the disc. We further use Au18 to demonstrate that the dipole pattern in the $\Delta v_\ell$ map can trace the chrono-kinematics of the inner Galaxy. 

 We compute the angle and the strength of the bar dipole signal after subtracting the background rotation based on the quasi-isothermal disc model \citep[see][]{ZS23}. We show that amongst the Galactic Mira variables, there are no obvious bar signatures in the period bin of $150-190$ days, drawing the same conclusion from the visual inspection of the residual $v_\ell$ map as well. This result suggests that the O-rich Mira variables with a period shorter than 190 days are born before the Milky Way's spin-up. This period corresponds to an age of $10.5\pm1.2$~Gyr or $9.1\pm1.0$~Gyr depending on the choice of the period-age relation and assuming a $11\%$ age uncertainty~\citep{ZS23}. We constrain the Milky Way disc to be younger than $11-12$ Gyr. Our results are consistent the previous measurements of the Milky Way spin-up time either from metallicity, stellar ages or globular cluster ages \citep{Miglio_2021, Queiroz_2023, Gallart_2024, Belokurov_2024}. We also discuss the role of the age uncertainty in the measurement of the disc formation epoch and show that age uncertainty leads to an overestimation of the disc age. In the future, we advocate using forward modelling when measuring the disc formation epoch, taking the age uncertainties fully into account. 

\section*{Data availability}

All data used in this work is publicly available. 

\section*{Acknowledgements}

We thank the reviewer for the helpful comments.
We thank Sergey Koposov for suggesting the test of the Fourier dipole moment. We thank Keith Hawkins, Sarah Kane, Adam Dillamore, Elliot Davies, Juntai Shen, and Chengye Cao for their inspirational discussions. 

HZ thanks the Science and Technology Facilities Council (STFC) for a PhD studentship. 
VB acknowledges support from the Leverhulme Research Project Grant RPG-2021-205: "The Faint Universe Made Visible with Machine Learning". 
JLS acknowledges support from the Royal Society (URF\textbackslash R1\textbackslash191555).
AAA acknowledges support from the Herchel Smith Fellowship at the University of Cambridge and a Fitzwilliam College research fellowship supported by the Isaac Newton Trust.

This work has made use of data from the European Space Agency (ESA) mission
{\textit{Gaia}} (\url{https://www.cosmos.esa.int/gaia}), processed by the {\textit{Gaia}}
Data Processing and Analysis Consortium (DPAC,
\url{https://www.cosmos.esa.int/web/gaia/dpac/consortium}). Funding for the DPAC
has been provided by national institutions, in particular the institutions
participating in the {\textit{Gaia}} Multilateral Agreement.





\bibliographystyle{mnras}
\bibliography{bibliography} 








\appendix

\section{Effects from distance uncertainty}
\label{Appendix::distance_uncertainty}

\begin{figure}
    \centering
    \includegraphics[width=\columnwidth]{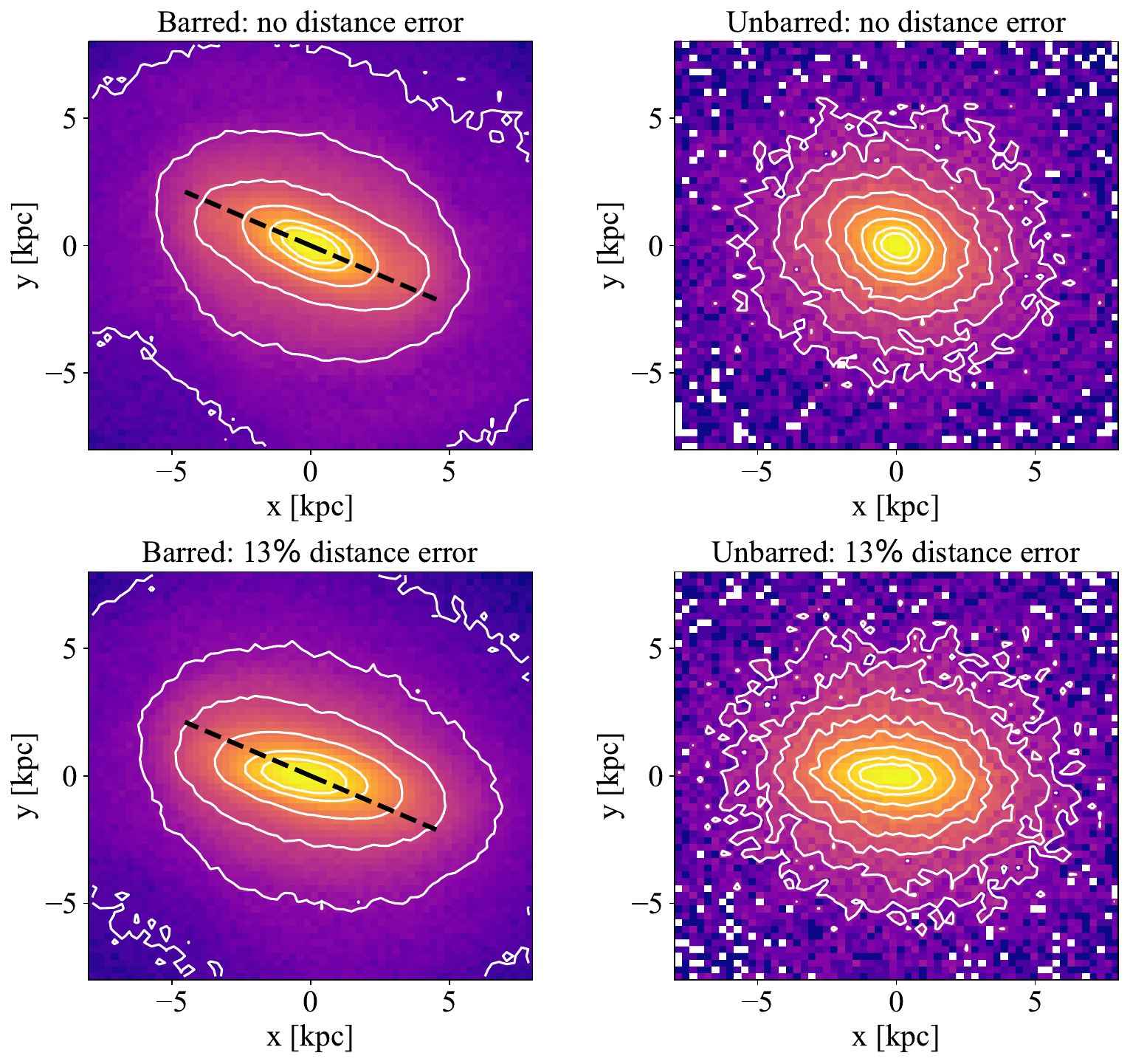}
    \caption{The effect of the heliocentric distance uncertainty on the observed surface density distribution. 
    \textit{Top left:} the surface density of a barred population without distance uncertainty. \textit{Bottom left:} the surface density of a barred population with a $13\%$ distance uncertainty. \textit{Top right:} the surface density of an unbarred population without distance uncertainty. \textit{Top right:} the surface density of an unbarred population with a $13\%$ distance uncertainty. The black dashed line in the left panels is used to label the true orientation of the Galactic bar.}
    
    \label{fig::appdix::dist_err}
\end{figure}

It has been shown that the heliocentric distance uncertainty affects the kinematic maps of the Galactic bar by biasing the bar angle towards the Sun-GC line \citep{Hey_2023, Vislosky_2024, Zhang_2024}. This is because the distance uncertainty smears the stars along the line-of-sight, and due to the geometry between the location of the Sun and the Galactic bar, the Galactic bar is sheared almost horizontally. As a result, the Galactic bar appears to have a smaller inclination angle with a larger distance uncertainty. We demonstrate this on the stellar surface density map. For convenience, we use Au18 but separate stars according to age to investigate the effects of distance uncertainty on both barred and unbarred populations. In Au18, stars with $\mathrm{age}<12$ Gyr are barred, and stars with $\mathrm{age}>12$ Gyr are unbarred.

First, we show the surface density map of the barred and unbarred populations without distance uncertainty in the top panels of Fig.~\ref{fig::appdix::dist_err}. The black dashed line shows the true orientation of the Galactic bar. Then, we scatter the stars with the typical distance uncertainty in the Mira sample we constructed, i.e. $13\%$. The surface density distribution of stars with added distance uncertainty is shown in the bottom panels. Comparing the bottom panel to the top panel for the barred population, we see the orientation of the Galactic bar is flattened towards the Sun-GC line (x-axis) due to the horizontal shear, as discussed above. For the unbarred population, it shows a high central concentration originally without distance uncertainty. However, the distance uncertainty stretches the centrally concentrated stars into a bar-like overdensity that aligns with the Sun-GC line, as shown in the lower right panel. The distance uncertainty-created, bar-like overdensity has a minor-to-major axis ratio of $\sim0.4$ similar to that of the real Galactic bar \citep{Bovy_2019}.

We demonstrated in Section~\ref{sec::spin_up_epoch_MW} that no bar signatures are found from the kinematics of the inner Galaxy for Mira variables with a period of 150-190 days. However, a bar-like overdensity that aligns with the $x$-axis is seen for those Mira variables, as shown in the top-right panel of Fig.~\ref{fig::Mira_bin}, which is also observed in the short-period Mira sample in \citet{Grady_2020} with a similar minor-to-major axis ratio of $\sim0.4$. According to the analysis of the distance uncertainty effects, we conclude this bar-like overdensity with zero inclination is likely induced by the imperfect distance uncertainty.

\bsp	
\label{lastpage}
\end{document}